\documentclass[printer]{aa}   
\usepackage{graphicx} 
\usepackage{txfonts} 
\usepackage{longtable,lscape} 
\usepackage{natbib} 
 
\begin{document} 
   \title{Structural properties of disk galaxies}    
 
   \subtitle{I. The intrinsic equatorial ellipticity of bulges} 
 
   \author{J. M\'endez-Abreu 
          \inst{1,2,4} 
          \and 
          J. A. L Aguerri \inst{3}
	  \and 
	  E. M. Corsini
	  \inst{4} 
	  \and  
	  E. Simonneau 
	  \inst{5} 
          } 
 
   \offprints{J. M\'endez-Abreu} 
 
   \institute{INAF-Osservatorio Astronomico di Padova, vicolo 
         dell'Osservatorio~5, I-35122 Padova, Italy\\ 
         \email{jairo.mendez@oapd.inaf.it}  
         \and Universidad de La 
         Laguna, Av. Astrof\'isico Francisco S\'anchez s/n, E-38206 La 
         Laguna, Spain  
         \and Instituto Astrof\'\i sico de 
         Canarias, Calle V\'ia L\'actea s/n, E-38200 La Laguna, Spain\\ 
         \email{jalfonso@iac.es}  
         \and Dipartimento di Astronomia, 
         Universit\`a di Padova, vicolo dell'Osservatorio~3, I-35122 
         Padova, Italy\\  
         \email{enricomaria.corsini@unipd.it}  
         \and 
         Institut d'Astrophysique de Paris, C.N.R.S., 98bis Boul. Arago, F-75014 
         Paris, France} 
 
   \date{\today} 
  \abstract  
  {A variety of formation scenarios have been proposed to explain the
  diversity of properties observed in bulges. Studying their intrinsic
  shape can help to constraint the dominant mechanisms at the epochs
  of their assembly.}
  {The structural parameters of a magnitude-limited sample of 148
  unbarred S0--Sb galaxies were derived in order to study the
  correlations between bulges and disks, as well as the probability
  distribution function of the intrinsic equatorial ellipticity of
  bulges.}
  {We present a new fitting algorithm (GASP2D) to perform
  two-dimensional photometric decomposition of the galaxy
  surface-brightness distribution. This was assumed to be the sum of
  the contribution of a bulge and disk component characterized by
  elliptical and concentric isophotes with constant (but possibly
  different) ellipticity and position angles. Bulge and disk
  parameters of the sample galaxies were derived from the $J-$band
  images, which were available in the Two Micron All Sky Survey. The
  probability distribution function of the equatorial ellipticity of
  the bulges was derived from the distribution of the observed
  ellipticities of bulges and misalignments between bulges and disks.}
  {Strong correlations between the bulge and disk parameters were
  found.  About $80\%$ of bulges in unbarred lenticular and
  early-to-intermediate spiral galaxies are not oblate but triaxial
  ellipsoids. Their mean axial ratio in the equatorial plane is
  $\langle B/A \rangle = 0.85$. Their probability distribution
  function is not significantly dependent on morphology, light
  concentration or luminosity. The possible presence of nuclear bars
  does not influence our results.}
  {The interplay between bulge and disk parameters favors scenarios in
  which bulges have assembled from mergers and/or have grown over long
  times through disk secular evolution. However, all these mechanisms
  have to be tested against the derived distribution of bulge
  intrinsic ellipticities.}
 
 
   \maketitle 
 
\section{Introduction}  
\label{sec:introduction} 
 
The relative prominence of galactic bulges with respect to their disks
is important in the definition of galaxy types. Therefore,
understanding the formation of bulges is key to understanding the
origin of the Hubble sequence.
 
Bulges are diverse and heterogeneous \citep[see the reviews 
by][]{kormendy93,wyse97,kormendy04}. 
The big bulges of lenticulars and early-type spirals are similar to
low-luminosity elliptical galaxies. Their surface-brightness radial
profiles generally follow a De Vaucouleurs law \citep[][hereafter
MH01]{andredakis95,carollo98,mollenhoff01}. The majority of these
bulges appear rounder than their associated disks
\citep{kent85}. Their kinematical properties are well described by
dynamical models of rotationally flattened oblate spheroids with
little or no anisotropy \citep{kormendy82,davies83,cappellari06}. They
have photometrical and kinematical properties, which satisfy the
fundamental plane (FP) correlation
\citep{bender92,bender93,burstein97,aguerri05a}.
On the contrary, the small bulges of late-type spiral galaxies seems
to be reminiscent of disks. Their surface-brightness radial profiles
have an almost exponential falloff
\citep{andredakis94,dejong96,macarthur03}. In some cases they have
apparent flattenings that are similar or even larger than their
associated disks \citep{fathi03} and rotate as fast as disks
\citep{kormendy93,kormendy02}. Late-type bulges deviate from the FP
\citep{carollo99}.
 
Different formation mechanisms (or at least a variety of dominant
mechanisms at the epochs of star formation and mass assembly) were
proposed to explain the variety of properties observed in bulges.
Some of these formation processes are rapid. They include early
formation from the dissipative collapse of protogalactic gas clouds
\citep{eggen62,sandage90,gilmore98,merlinchiosi06} or later assembly
from mergers between pre-existing disks
\citep{kauffmann96,baugh96,cole00}. In both scenarios the disk forms
after the bulge as a consequence of either a long star-formation time
compared to the collapse time or a re-accretion around the newly
formed bulge.
 
Bulges can also grow over long timescales through the disk secular
evolution driven by bars and/or environmental effects.
Bars are present in more than half of disk galaxies in the local
universe \citep{eskridge00,menendez-delmestre07} and out to $z\sim1$
\citep{elmegreen04,jogee04}. They are efficient mechanisms for driving
gas inward to the galactic center and feed the galactic supermassive
black hole \citep[see][and references therein]{corsini03a}. In
addition, bar dissolution due to the growth of a central mass
\citep{pfenninger90}, scattering of disk stars at vertical resonances
\citep{combes90}, and coherent bending of the bar perpendicular to the
disk plane
\citep{raha91,debattista04,athanassoula05,martinez-valpuesta06} are
efficient mechanisms in building central bulge-like structures, the
so-called boxy/peanut bulges.
Moreover, the growth of the bulge out of disk material may also be 
externally triggered by satellite accretion during minor merging 
events \citep{searlezinn78,aguerri01,eliche-moral06} and gas infall 
\citep{thakarryden98}. 
 
Traditionally, the study of the relations between the structural
parameters of the galaxies have been used to understand the bulge
formation processes, e.g., the correlation between the bulge effective
radius and the scale length of the disk in many galaxy samples has
always been interpreted as an indication that bulges were formed by
secular evolution of their disks \citep[see][]{macarthur03}. However,
one piece lost in this study is the three-dimensional shape of the
bulges.
By studying this, one might be able to provide the relative importance
of rapid and slow processes in assembling the dense central components
of disk galaxies.
A statistical study can provide a crucial piece of information for 
testing the results of numerical simulations of bulge formation for 
different galaxy type along the morphological sequence. 
 
In this paper, we analyze a sample of unbarred early-type disk
galaxies to derive the intrinsic ellipticity of their bulges in the
galactic plane.
The twisting of bulge isophotes \citep{lindblad56,zaritsky86} and
misalignment between the major axes of the bulge and disk
\citep{bertola91} are not possible if the bulge and disk are both
oblate.  Therefore, they were interpreted as a signature of bulge
triaxiality.  This idea is supported by the presence of non-circular
gas motions
\citep[e.g.,][]{gerhardvietri86,bertola89,gerhard89,berman01} and a
velocity gradient along the galaxy minor axis
\citep{corsini03b,coccato04,coccato05}.
We improve the previous works in several aspects. First, we use
near-infrared images to map the distribution of the mass-carrying
evolved stars and avoid contamination of dust and bright young stars.
Second, we retrieve the structural parameters of the bulge and disk by
applying a new algorithm for two-dimensional photometric decomposition
of the observed surface-brightness distribution.  Finally, we obtain
the probability distribution function (PDF) of the intrinsic
equatorial ellipticity of bulges by using a new mathematical treatment
of the equations describing their three-dimensional shape.
 
The paper is organized as follow. The selection criteria of our sample
galaxies and the analysis of their near-infrared images are described
in Sect. \ref{sec:sample}. Our new photometric decomposition method
for deriving the structural parameters of the bulge and disk by
analyzing the two-dimensional surface brightness distribution of
galaxies is presented in Sect. \ref{sec:decomposition}. The
correlations between the structural parameters of the sample galaxies
are discussed in Sect. \ref{sec:correlations}. The PDF of intrinsic
equatorial ellipticity of the studied bulges is derived in
Sect. \ref{sec:ellipticity}. Our conclusions and a summary of the
results are given in Sect. \ref{sec:discussion}.
 
\section{Sample selection and data acquisition} 
\label{sec:sample} 
 
Our objective was to select a well-defined complete sample of nearby 
unbarred disk galaxies to study in a systematic way the photometric 
properties of their structural components. 
Since these properties are strongly dependent on the dominating 
stellar population at the observed wavelength, it is preferable to 
consider near-infrared images to map the mass-carrying evolved stars 
and avoid contamination due to dust and bright young stars 
(e.g., MH01). 
The complete sample is drawn from the Extended Source Catalogue (XSC)
\citep{Jarrett00} of the Two Micron All Sky Survey (2MASS)
\citep{skrutskie06}. Our sample consists of galaxies that meet the
following requirements:
(1) Hubble type classification from S0 to Sb ($\rm -3 \leq HT \leq 3$; 
    de Vaucouleurs et al. 1991, hereafter RC3) to ensure that bulges 
    are fully resolved in 2MASS images; 
(2) unbarred classification in RC3; 
(3) total $J-$band magnitude $J_T<10$ mag (2MASS/XSC); 
(4) inclination $i<65^{\circ}$ (RC3) to measure the misalignment
    between the position angle of the bulge and disk;
(5) Galactic latitude $\mid b_{G}\mid > 30^{\circ}$ (RC3) to minimize 
    both Galactic extinction and contamination due to Galactic 
    foreground stars. 
 
We ended up with a sample of 184 bona-fide unbarred galaxies.   
We retrieved their 2MASS $J-$band images from the NASA/IPAC Infrared 
Science Archive. The galaxy images were reduced and flux calibrated 
with the standard 2MASS extended source processor GALWORKS 
\citep{Jarrett00}. Images have a typical field of view of few 
arc-minutes and a spatial scale of $1''$ pixel$^{-1}$. They were 
obtained with an average seeing ${\it FWHM}$$\sim$3$\farcs1$ as 
measured by fitting a circular two-dimensional Gaussian to the field 
stars. 
 
\begin{figure} 
\centering 
\includegraphics[width=9cm]{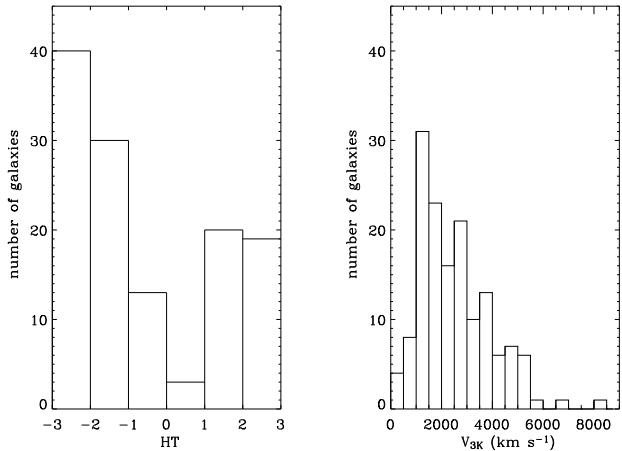} 
\caption{Distribution of the sample galaxies over the Hubble 
  types (left panel) and radial velocities with respect to the 
  3K background (right panel).} 
\label{fig:hist_hubble} 
\end{figure}

After a visual inspection of the images, we realized that some of the
sample galaxies were not suitable for our study. We rejected paired
and interacting objects as well as those galaxies that resulted in
being barred after performing the photometric decomposition (see
Sect. \ref{sec:decomposition}).  Therefore, the final sample presented
in this paper contains 148 galaxies (90 lenticular and 58 early-type
spiral galaxies). A compilation of their main properties is given in
Table \ref{tab:sample}.
Figure \ref{fig:hist_hubble} shows the distribution of the sample
galaxies over the Hubble types. The lenticular galaxies are
predominant over the spirals due to our magnitude selection, which
favors red galaxies. Moreover, we show the distribution of radial
velocities of the sample galaxies with respect to the 3K
background. The mean radial velocity is 2000 km~s$^{-1}$
(corresponding to a distance of 27 Mpc by assuming $H_0=75$
km~s$^{-1}$~Mpc$^{-1}$), but we include galaxies as far as 8500
km~s$^{-1}$ (113 Mpc) because the sample is magnitude limited.

Tonry et al. (2001) derived the distance of 30 galaxies of our sample
from the measurement of their surface brightness fluctuations.  The
difference between the distances obtained from radial velocities and
those derived from surface brightness fluctuations was calculated for
all these galaxies. The standard deviation of the distance differences
is 5 Mpc and it was assumed as being a typical distance error. For the
4 common galaxies in the Virgo cluster it is 2 Mpc.

\section{Two-dimensional bulge-disk parametric decomposition} 
\label{sec:decomposition} 
 
Conventional bulge-disk decompositions based on elliptically averaged
surface-brightness profiles usually do not take into account the
intrinsic shapes \citep[e.g.,][]{prieto01} or the position angle
\citep[e.g.,][]{trujillo01c} of the bulge and disk components, which
can produce systematic errors in the results \citep[e.g.,][]{byun95}.
 
For this reason a number of two-dimensional parametric decomposition 
techniques have been developed in the last years. As examples we may 
point out the algorithms developed by Simard (1998, GIM2D), Peng et 
al. (2002, GALFIT), de Souza et al. (2004, BUDDA) and Pignatelli et 
al. (2006, GASPHOT). These methods were developed to solve different 
problems of galaxy decomposition when fitting the two-dimensional 
galaxy surface-brightness distribution. They use different functions 
to parametrize the galaxy component and different minimizations 
routines to perform the fit. 
 
In this paper we present our new decomposition algorithm named GASP2D 
(GAlaxy Surface Photometry 2 Dimensional Decomposition). The code 
works like GIM2D and GASPHOT in minimizing the interaction with 
the user. It works in an automatical way to be more efficient when 
dealing with a large amount of galaxies. However, like GALFIT and 
BUDDA it also adopts a Leverberg-Marquard algorithm to fit the 
two-dimensional surface-brightness distribution of the galaxy. This 
reduces the amount of computational time needed to obtain a robust and 
reliable estimate of the galaxy structural parameters. 
 
In the present work, we show the first version of the code. We assume
that the galaxy can be modeled with only two components, the bulge and
the disk. In a forthcoming paper, we will show an improved version of
GASP2D with the possibility to fit other galaxy components, like bars.
 
\subsection{Photometric model}  
\label{sec:model} 
 
We assumed the galaxy surface-brightness distribution to be the sum of 
the contribution of a bulge and disk component. Both of them are 
characterized by elliptical and concentric isophotes with constant 
(but possibly different) ellipticity and position angles. 
 
Let $(\xi$, $\eta$, $\zeta)$ be the Cartesian coordinates with the origin 
in the galaxy center, the $\xi-$axis parallel to the direction of right 
ascension and pointing westward, the $\eta-$axis parallel to the direction 
of declination and pointing northward, and the $\zeta-$axis along the 
line-of-sight and pointing toward the observer. The plane of the sky 
is confined to the $(\xi,\,\eta)$ plane. 
 
We adopted the S\'ersic law \citep{sersic68} to describe the surface 
brightness of the bulge component. The S\'ersic law has been 
extensively used in the literature for modeling the surface 
brightness profiles of galaxies. For instance, it has been used to 
model the surface brightness of elliptical galaxies 
\citep{grahamguzman03}, bulges of early and late type galaxies 
\citep{andredakis95,prieto01,aguerri04,mollenhoff04}, the low surface 
brightness host galaxy of blue compact galaxies \citep{caon05,amorin07} 
and dwarf elliptical galaxies 
\citep{binggeli98,aguerri05b,grahamguzman03}. It is given by
 
\begin{equation} 
I_{\rm b}(\xi,\eta)=I_{\rm e}10^{-b_n\left[\left(r_{\rm b}/r_{\rm e} 
\right)^{1/n}-1\right]},
\label{eqn:bulge_surfbright} 
\end{equation} 
%
where $r_{\rm e}$, $I_{\rm e}$, and $n$ are the effective (or 
half-light) radius, the surface brightness at $r_{\rm e}$, and a shape 
parameter describing the curvature of the surface-brightness profile, 
respectively. The value of $b_n$ is coupled to $n$ so that half of the 
total luminosity of the bulge is within $r_{\rm e}$ and can be 
approximated as $b_{n} = 0.868\,n - 0.142$ \citep{caon93}. 
Bulge isophotes are ellipses centered on $(\xi_0,\,\eta_0)$ with constant 
position angle PA$_{\rm b}$ and constant ellipticity $\epsilon_{\rm b} 
= 1 - q_{\rm b}$. 
The radius $r_{\rm b}$ is given by 
 
\begin{eqnarray} 
r_{\rm b} & = &\left[(-(\xi-\xi_0)\sin{{\rm PA_b}} + (\eta-\eta_0)  
\cos{{\rm PA_b}})^2 + \right. \nonumber \\  
    &   & - \left.((\xi-\xi_0)\cos{{\rm PA_b}} + (\eta-\eta_0)  
\sin{{\rm PA_b}})^2/q_{\rm b}^2\right]^{1/2}.  
\label{eqn:bulge_radius} 
\end{eqnarray} 
 
We adopted the exponential law \citep{freeman70} to describe the 
surface brightness of the disk component 
 
\begin{equation} 
I_{\rm d}(\xi,\eta) = I_0\,e^{-r_{\rm d}/h}, 
\label{eqn:disc_surfbright} 
\end{equation} 
%
where $I_0$ and $h$ are the central surface brightness and scalelength 
of the disk, respectively. 
Disk isophotes are ellipses centered on $(\xi_0,\,\eta_0)$ with 
constant position angle PA$_{\rm d}$ and constant ellipticity 
$\epsilon_{\rm d} = 1 - q_{\rm d}$. Disk inclination is $i = 
\arccos{(q_{\rm d})}$. 
The radius $r_{\rm d}$ is given by 
 
\begin{eqnarray} 
r_{\rm d} & = &\left[(-(\xi-\xi_0) \sin{{\rm PA_d}} + (\eta-\eta_0)  
\cos{{\rm PA_d}})^2 + \right. \nonumber \\  
    &   & - \left.((\xi-\xi_0) \cos{{\rm PA_d}} + (\eta-\eta_0)  
\sin{{\rm PA_d}})^2/q_d^2\right]^{1/2}. 
\label{eqn:disc_radius} 
\end{eqnarray} 
 
To derive the coordinates $(\xi_0,\,\eta_0)$ of the galaxy center and
the photometric parameters of the bulge ($I_{\rm e}$, $r_{\rm e}$,
$n$, PA$_{\rm b}$, and $q_{\rm b}$) and disk ($I_{0}$, $h$, PA$_{\rm
d}$, and $q_{\rm d}$) we fitted iteratively a model of the surface
brightness $I_{\rm m}(\xi,\,\eta)$=$I_{\rm b}(\xi,\,\eta)$+$I_{\rm
d}(\xi,\,\eta)$ to the observations using a non-linear least-squares
minimization method. It was based on the robust Levenberg-Marquardt
method \citep[e.g.,][]{press96} implemented by More et al. (1980). The
actual computation was done using the MPFIT algorithm implemented by
C.~B. Markwardt under the IDL environment\footnote{The updated version
of this code is available on
http://cow.physics.wisc.edu/~craigm/idl/idl.html}. MPFIT allows the
user to keep constant or impose boundary constraints on any parameter
during the fitting process.
 
For each pixel $(\xi,\,\eta)$, the observed galaxy photon counts
$I_{\rm g}(\xi,\,\eta)$ are compared with those predicted from the
model $I_{\rm m}(\xi,\,\eta)$. Each pixel is weighted according to the
variance of its total observed photon counts due to the contribution
of both galaxy and sky, and determined assuming photon noise
limitation by taking into account the detector readout noise
(RON). Therefore, the $\chi^2$ to be minimized can be written as
 
\begin{equation}     
  \chi^{2} = \sum_{\xi=1}^{N} \sum_{\eta=1}^{M}        
  \frac{[\,I_{\rm m}\,(\xi,\,\eta) - I_{\rm g}\,(\xi,\,\eta)\,]^2}{
  I_{\rm g}\,(\xi,\,\eta) +I_{\rm s}\,(\xi,\,\eta)+ {\rm RON}^2}, 
\label{eqn:chi2}     
\end{equation}     
%
with $\xi$ and $\eta$ ranging over the whole $N \times M$ pixel image. 
 
An important point to consider here is the weight function used to
calculate the $\chi^2$; some authors claim that is better to assign to
each pixel an uncertainty given by the Poissonian noise
\cite[e.g.,][]{peng02} while others adopt constant weights to obtain
better results (e.g., MH01; de Souza et al. 2004). Both possibilities
were implemented in the fitting algorithm. We adopted the poissonian
weights after extensive testing with artificial galaxies.
  
\subsection{Seeing effects} 
\label{sec:seeing} 
 
The ground-based images are affected by seeing, which scatters the 
light of the objects and produces a loss of spatial resolution. This 
is particularly critical in the central regions of galaxies, where the 
slope of the radial surface brightness profile is steeper. Since the 
bulge contribution dominates the surface brightness distribution at 
small radii, seeing mostly affects bulge structural parameters. Seeing 
effects on the scale parameters of a S\'ersic surface brightness 
profile have been extensively discussed by Trujillo et al. (2001a,b). 
 
During each iteration of the fitting algorithm, the seeing effects
were taken into account by convolving the model image with a circular
two-dimensional Gaussian point spread function (PSF). The PSF {\em
FWHM} was chosen to match the one measured from the foreground stars
in the field of the 2MASS galaxy image. The convolution was performed
in the Fourier domain using the fast Fourier transform (FFT) algorithm
\citep{press96} before the $\chi^2$ calculation.
Our code allows us to introduce also a Moffat or a star image to
reproduce the PSF. We tested that the results are not improved by
adopting a circular two-dimensional Moffat PSF or by computing the
convolution integrals.

\subsection{Technical procedure of the fit} 
\label{sec:trials} 
 
Since the fitting algorithm is based on a $\chi^2$ minimization, it is 
important to adopt initial trials for free parameters as close as 
possible to their actual values. This would ensure that the iteration 
procedure does not just stop on a local minimum of the $\chi^2$ 
distribution. To this aim we proceeded through different steps. 
 
First, the photometric package SExtractor \cite[see][for
details]{bertinarnouts96} was used to measure position, magnitude and
ellipticity of the sources (e.g., foreground stars, background and
companion galaxies, as well as bad pixels) in the images.
 
We then derived the elliptically-averaged radial profiles of the
surface brightness, ellipticity and position angle of the galaxy. We
fitted ellipses to the galaxy isophotes with the ELLIPSE task in
IRAF\footnote{IRAF is distributed by NOAO, which is operated by AURA
Inc., under contract with the National Science Foundation.}. After
masking the spurious sources using the parameters provided by
SExtractor, we fitted ellipses centered on the position of the galaxy
center. This could be estimated by either a visual inspection of the
image or with SExtractor. The coordinates of the galaxy center were
adopted as initial trials for $(\xi_0,\,\eta_0)$ in the
two-dimensional fit.
 
In the third step we derived some of the trial values by performing a 
standard one dimensional decomposition technique similar to that 
adopted by several authors \citep[e.g.,][]{kormendy77,prieto01}. 
We began by fitting an exponential law to the radial 
surface-brightness profile at large radii, where the light 
distribution of the galaxy is expected to be dominated by the disk 
contribution. The central surface brightness and scalelength of the 
fitted exponential profile were adopted as initial trials for $I_0$ 
and $h$, respectively. 
The fitted profile was extrapolated at small radii and then subtracted 
from the observed radial surface-brightness profile. The residual 
radial surface-brightness profile was assumed to be a first estimate 
of the light distribution of the bulge. We fitted it with a S\'ersic 
law by assuming the bulge shape parameter to be $0.5, 1, 1.5,..., 6$. 
The bulge effective radius, effective surface brightness, and shape
parameter that (together with the disk parameters) gave the best fit
to the radial surface-brightness profile were adopted as initial
trials for $r_{\rm e}$, $I_{\rm e}$, and $n$, respectively.
  
We also obtained the initial trials for ellipticity and position angle
of the disk and bulge, respectively. The values for $q_{\rm d}$ and
PA$_{\rm d}$ were found by averaging the values in the outermost
portion of the radial profiles of ellipticity and position angle.
The initial trials for $q_{\rm b}$ and PA$_{\rm b}$ were estimated by 
interpolating at $r_{\rm e}$ the radial profiles of the ellipticity 
and position angle, respectively. 
 
Finally, the initial guesses were adopted to initialize the non-linear 
least-squares fit to galaxy image, where all the parameters, including 
$n$, were allowed to vary. A convergent model was reached when the 
$\chi^2$ had a minimum and the relative change of the $\chi^2$ between 
the iterations was less than $10^{-7}$. 
A model of the galaxy surface-brightness distribution was built using
the fitted parameters. It was convolved with the adopted circular
two-dimensional Gaussian PSF and subtracted from the observed image to
obtain a residual image. In order to ensure the minimum in the
$\chi^2$-space found in this first iteration, we perform two more
iterations. In these iterations, all the pixels and/or regions of the
residual image with values greater or less than a fixed threshold,
controlled by the user, were rejected. Those regions were masked out
and the fit was repeated assuming, as initial trials for the free
parameters, the values obtained in the previous iteration. These kind
of masks are usually useful when galaxies have other prominent
structures, different from the bulge and disk (e.g. spiral arms and
dust lanes), which can affect the fitted parameters, solving the
problem in an automatic way. We found that after three iterations the
algorithm converges and the parameters do not change.
 
\subsection{Test on simulated galaxies} 
\label{sec:test} 
 
To test the reliability and accuracy of our two-dimensional technique
for bulge-disk decomposition, we carried out extensive simulations on
a large set of artificial disk galaxies. We generated 1000 images of
galaxies with a S\'ersic bulge and an exponential disk. The central
surface-brightness, scalelength, and apparent axial ratios of the
bulge and disk of the artificial galaxies were randomly chosen in the
range of values observed in the $J-$band by MH01 for a sample of 40
bright spiral galaxies. The adopted ranges were
 
\begin{equation}     
1 \leq r_{\rm e} \leq 3\ {\rm kpc} \qquad 0.4 \leq q_{\rm b} \leq 0.9 
  \qquad 0.5 \leq n \leq 6\\    
\label{eqn:re_limits} 
\end{equation}     
    
\noindent     
for the bulge parameters, and     
         
\begin{equation}     
2 \leq h \leq 5\ {\rm kpc} \qquad 0.4 \leq q_{\rm d} \leq 0.9 \\     
\end{equation}     
%
for the disk parameters. The parameters of the artificial galaxies
also have to satisfy the following conditions
 
\begin{equation}     
q_{\rm d} < q_{\rm b} \qquad 0 < B/T < 0.8 \qquad 8 < J_T < 12 \ {\rm mag}. 
\end{equation}     
%
 
All the simulated galaxies were assumed to be at a distance of $30$
Mpc. This is the average distance of the galaxies of our sample and it
corresponds to a scale of $145$ pc arcsec$^{-1}$.  The pixel scale
used was 1 arcsec pixel$^{-1}$, and the CCD gain and RON were 8
e$^{-}$ ADU$^{-1}$ and 40 e$^{-}$, respectively, in order to mimic the
instrumental setup of the 2MASS data. Finally, a background level and
photon noise were added to the resulting images to yield a
signal-to-noise ratio similar to that of the available 2MASS images.
 
The two-dimensional parametric decomposition was applied to analyze 
the images of the artificial galaxies as if they were real. Errors on 
the fitted parameters were estimated by comparing the input $p_{\rm 
i}$ and output $p_{\rm o}$ values. Relative errors ($1-p_{\rm 
i}/p_{\rm o}$) were assumed to be normally distributed, with mean and 
standard deviation corresponding to the systematic and typical error 
on the relevant parameter, respectively.  

The results of the simulations are given in
Table~\ref{tab:simulations}.
In Run 1 we built the artificial galaxies by assuming the correct
values of PSF {\em FWHM} and sky level, so only errors due to the
Poisson noise are studied. The mean relative errors on the fitted
parameters are smaller than $0.01$ (absolute value) and their standard
deviations are smaller than $0.02$ (absolute value) for all galaxies
with $J_T<10$ mag, proving the reliability of our derived structural
parameters. Relative errors increase for fainter galaxies, which are
not included in our sample.

Systematic errors given by a wrong estimation of PSF and sky level are 
the most significant contributors to the error budget.   
To understand how a typical error in the measurement of the PSF {\em
FWHM} affects our results, we analyzed the artificial galaxies by
adopting the correct sky level and a PSF {\em FWHM} that was $2\%$
larger (Run 2) or smaller (Run 3) than the actual one. As expected
(Sect. \ref{sec:seeing}), the parameters of the surface-brightness
profiles show larger errors for the bulge than for the disk. We
recovered larger values for the S\'ersic parameters (r$_{e}$, $n$)
when the PSF {\em FWHM} is overestimated, and lower values when it is
underestimated. Relative errors are correlated with the values of
effective radius and shape parameter of the bulge but not with the
magnitude of the galaxy.
In Run 4 we built the artificial galaxies by adopting the correct PSF
{\em FWHM} and a sky level that was $1\%$ larger than the actual
one. For brighter galaxies an improper sky subtraction mostly affects
the parameters of the disk surface-brightness profile. For fainter
galaxies, the large relative errors on the bulge parameters are due to
their coupling with the disk parameters. This is consistent with the
results of similar tests performed by Byun \& Freeman (1995).
 
The structural parameters to be measured to derive the intrinsic shape
of bulges are the ellipticity of the bulge and position angles of the
bulge and disk (Sect. \ref{sec:deprojection}). In all the runs, the
relative errors are smaller than $0.05$ (absolute value) for galaxies
with $J_T<10$ mag. Larger errors ( up to about 0.1) were found for
fainter galaxies after an improper subtraction of the sky level from
the image. However, this is not the case for our sample galaxies.
 
\begin{table*}
\begin{scriptsize}
\begin {center}
  \caption{Relative errors on the photometric parameters of the bulge
  and disk calculated for different galaxy magnitudes by means of
  Monte Carlo simulations.}
\label{tab:simulations}
\begin{tabular}{lccrccrccrccrc}
\hline
\noalign{\smallskip}
\multicolumn{2}{c}{} &
\multicolumn{3}{c}{$8 < J_{T} \leq 9$} &
\multicolumn{3}{c}{$9 < J_{T} \leq 10$} &
\multicolumn{3}{c}{$10 < J_{T} \leq 11$} &
\multicolumn{3}{c}{$11 < J_{T} \leq 12$} \\
\multicolumn{1}{l}{Parameter} &
\multicolumn{1}{c}{Run} &
\multicolumn{1}{c}{\#} &
\multicolumn{1}{c}{Mean} &
\multicolumn{1}{c}{St.Dev.} &
\multicolumn{1}{c}{\#} &
\multicolumn{1}{c}{Mean} &
\multicolumn{1}{c}{St.Dev.} &
\multicolumn{1}{c}{\#} &
\multicolumn{1}{c}{Mean} &
\multicolumn{1}{c}{St.Dev.} &
\multicolumn{1}{c}{\#} &
\multicolumn{1}{c}{Mean} &
\multicolumn{1}{c}{St.Dev.} \\
\multicolumn{1}{l}{(1)} &
\multicolumn{1}{c}{(2)} &
\multicolumn{1}{c}{(3)} &
\multicolumn{1}{c}{(4)} &
\multicolumn{1}{c}{(5)} &
\multicolumn{1}{c}{(6)} &
\multicolumn{1}{c}{(7)} &
\multicolumn{1}{c}{(8)} &
\multicolumn{1}{c}{(9)} &
\multicolumn{1}{c}{(10)} &
\multicolumn{1}{c}{(11)} &
\multicolumn{1}{c}{(12)} &
\multicolumn{1}{c}{(13)} &
\multicolumn{1}{c}{(14)} \\
\noalign{\smallskip}
\hline
\noalign{\smallskip}
$I_{\rm e}$ & 1 & 226 & $-1\cdot10^{-3}$ & 0.02 & 217 &  $-8\cdot10^{-3}$ & 0.04 &
238 & $-1\cdot10^{-2}$ & 0.06 & 206 & $-4\cdot10^{-2}$ &0.14 \\
            & 2 & 223 & $1\cdot10^{-2}$  & 0.08 & 211 &  $1\cdot10^{-1}$  & 0.09 &
235 & $9\cdot10^{-2}$  & 0.10 & 205 & $2\cdot10^{-2}$  & 0.16 \\
            & 3 & 225 & $-1\cdot10^{-2}$ & 0.09 & 214 & $-1\cdot10^{-1}$  & 0.09 &
231 & $-8\cdot10^{-2}$ & 0.10 & 208 & $-6\cdot10^{-2}$ & 0.15 \\
            & 4 & 212 & $-6\cdot10^{-2}$ & 0.08 & 206 & $-7\cdot10^{-2}$  & 0.10 &
220 & $-6\cdot10^{-2}$ & 0.11 & 180 & $-3\cdot10^{-2}$ & 0.15 \\
\noalign{\smallskip}
\hline
\noalign{\smallskip}
$r_{\rm e}$ & 1 & 226 & $1\cdot10^{-3}$  & 0.02 & 217 &  $7\cdot10^{-3}$  & 0.03 &
238 & $1\cdot10^{-2}$  & 0.06 & 206 &  $6\cdot10^{-2}$ & 0.13 \\
            & 2 & 223 & $-7\cdot10^{-2}$ & 0.08 & 211 &$-9\cdot10^{-2}$   & 0.08 &
235 & $-7\cdot10^{-2}$ & 0.09 & 205 & $-2\cdot10^{-2}$ & 0.14 \\
            & 3 & 225 & $7\cdot10^{-2}$  & 0.06 & 214 &  $9\cdot10^{-2}$  & 0.07 &
231 & $1\cdot10^{-1}$  & 0.09 & 208 &  $1\cdot10^{-1}$ & 0.13 \\
            & 4 & 212 & $7\cdot10^{-2}$  & 0.09 & 206 &  $8\cdot10^{-2}$  & 0.10 &
220 & $1\cdot10^{-1}$  & 0.11 & 180 &  $8\cdot10^{-2}$ & 0.14 \\
\noalign{\smallskip}
\hline
\noalign{\smallskip}
$n$         & 1 & 226 & $2\cdot10^{-3}$  & 0.02 & 217 &  $4\cdot10^{-3}$  & 0.02 &
238 & $4\cdot10^{-3}$  & 0.04 & 206 &  $4\cdot10^{-2}$ & 0.09 \\
            & 2 & 223 & $-5\cdot10^{-2}$ & 0.05 & 211 & $-6\cdot10^{-2}$  & 0.05 &
235 & $-7\cdot10^{-2}$ & 0.07 & 205 & $-3\cdot10^{-2}$ & 0.12 \\
            & 3 & 225 & $5\cdot10^{-2}$  & 0.04 & 214 &  $6\cdot10^{-2}$  & 0.05 &
231 & $7\cdot10^{-2}$  & 0.07 & 208 &  $1\cdot10^{-1}$ & 0.11 \\
            & 4 & 212 & $1\cdot10^{-1}$  & 0.08 & 206 &  $1\cdot10^{-1}$  & 0.08 &
220 & $1\cdot10^{-1}$  & 0.09 & 180 &  $1\cdot10^{-1}$ & 0.15 \\
\noalign{\smallskip}
\hline
\noalign{\smallskip}
$I_{0}$     & 1 & 226 & $-3\cdot10^{-3}$ & 0.03 & 217 & $-6\cdot10^{-3}$  & 0.03 &
238 & $-8\cdot10^{-3}$ & 0.05 & 206 & $-3\cdot10^{-2}$ & 0.14 \\
            & 2 & 223 & $1\cdot10^{-2}$  & 0.05 & 211 & $1\cdot10^{-2}$   & 0.05 &
235 & $ 3\cdot10^{-2}$ & 0.07 & 205 & $2\cdot10^{-3}$  & 0.13 \\
            & 3 & 225 & $-2\cdot10^{-2}$ & 0.05 & 214 & $-3\cdot10^{-2}$  & 0.05 &
231 & $-3\cdot10^{-2}$ & 0.05 & 208 & $-3\cdot10^{-2}$ & 0.07 \\
            & 4 & 212 & $-7\cdot10^{-2}$ & 0.10 & 206 & $-9\cdot10^{-2}$  & 0.10 &
220 & $-9\cdot10^{-2}$ & 0.11 & 180 & $-1\cdot10^{-1}$ & 0.12 \\
\noalign{\smallskip}
\hline
\noalign{\smallskip}
$h$         & 1 & 226 & $2\cdot10^{-3}$  & 0.02 & 217 & $9\cdot10^{-3}$   & 0.03 &
238 & $9\cdot10^{-3}$  & 0.04 & 206 &  $4\cdot10^{-2}$ & 0.10 \\
            & 2 & 223 & $3\cdot10^{-3}$  & 0.02 & 211 &  $5\cdot10^{-3}$  & 0.03 &
235 & $6\cdot10^{-3}$  & 0.05 & 205 &  $2\cdot10^{-2}$ & 0.11 \\
            & 3 & 225 & $3\cdot10^{-3}$  & 0.02 & 214 &  $8\cdot10^{-3}$  & 0.03 &
231 & $1\cdot10^{-2}$  & 0.05 & 208 &  $6\cdot10^{-2}$ & 0.10 \\
            & 4 & 212 & $8\cdot10^{-2}$  & 0.09 & 206 &  $8\cdot10^{-2}$  & 0.08 &
220 & $9\cdot10^{-2}$  & 0.10 & 180 &  $1\cdot10^{-1}$ & 0.11 \\
\noalign{\smallskip}
\hline
\noalign{\smallskip}
$q_{\rm b}$ & 1 & 226 & $-1\cdot10^{-3}$ & 0.01 & 217 & $-4\cdot10^{-3}$  & 0.02 &
238 & $-2\cdot10^{-3}$ & 0.03 & 206 & $1\cdot10^{-2}$  & 0.06 \\
            & 2 & 223 & $3\cdot10^{-3}$  & 0.01 & 211 &  $3\cdot10^{-3}$  & 0.01 &
235 & $3\cdot10^{-3}$  & 0.02 & 205 &  $1\cdot10^{-2}$ & 0.06 \\
            & 3 & 225 & $-5\cdot10^{-3}$ & 0.01 & 214 & $-7\cdot10^{-3}$  & 0.02 &
231 & $-5\cdot10^{-3}$ & 0.03 & 208 & $-5\cdot10^{-3}$ & 0.06 \\
            & 4 & 212 & $1\cdot10^{-2}$  & 0.10 & 206 &  $4\cdot10^{-2}$  & 0.12 &
220 & $4\cdot10^{-2}$  & 0.12 & 180 &  $6\cdot10^{-2}$ & 0.13 \\
\noalign{\smallskip}
\hline
\noalign{\smallskip}
$q_{\rm d}$ & 1 & 226 & $-1\cdot10^{-3}$ & 0.01 & 217 & $-8\cdot10^{-3}$  & 0.04 &
238 & $-3\cdot10^{-3}$ & 0.03 & 206 & $-3\cdot10^{-2}$ & 0.09 \\
            & 2 & 223 & $8\cdot10^{-3}$  & 0.02 & 211 &  $8\cdot10^{-3}$  & 0.04 &
235 & $2\cdot10^{-2}$  & 0.04 & 205 &$2\cdot10^{-2}$   & 0.11 \\
            & 3 & 225 & $-1\cdot10^{-2}$ & 0.02 & 214 & $-2\cdot10^{-2}$  & 0.03 &
231 & $-2\cdot10^{-2}$ & 0.04 & 208 & $-5\cdot10^{-2}$ & 0.10 \\
            & 4 & 212 & $-8\cdot10^{-2}$ & 0.08 & 206 & $-7\cdot10^{-2}$  & 0.08 &
220 & $-9\cdot10^{-2}$ & 0.09 & 180 & $-1\cdot10^{-1}$ & 0.10 \\
\noalign{\smallskip}
\hline
\noalign{\smallskip}
PA$_{\rm b}$& 1 & 226 & $1\cdot10^{-4}$  & 0.01 & 217 &  $1\cdot10^{-3}$  & 0.02 &
238 & $-2\cdot10^{-3}$ & 0.02 & 206 & $-4\cdot10^{-3}$ & 0.06 \\
            & 2 & 223 & $-1\cdot10^{-3}$ & 0.02 & 211 &  $-1\cdot10^{-3}$ & 0.02 &
235 & $-2\cdot10^{-3}$ & 0.03 & 205 & $3\cdot10^{-3}$  & 0.05 \\
            & 3 & 225 & $1\cdot10^{-3}$  & 0.01 & 214 &  $3\cdot10^{-3}$  & 0.02 &
231 & $-3\cdot10^{-3}$ & 0.03 & 208 & $-1\cdot10^{-3}$ & 0.06 \\
            & 4 & 212 & $-1\cdot10^{-2}$ & 0.10 & 206 & $-2\cdot10^{-2}$  & 0.10 &
220 & $-3\cdot10^{-2}$ & 0.10 & 180 & $-1\cdot10^{-2}$ & 0.12 \\
\noalign{\smallskip}
\hline
\noalign{\smallskip}
PA$_{\rm d}$& 1 & 226 & $-3\cdot10^{-3}$ & 0.02 & 217 & $-3\cdot10^{-4}$  & 0.02 &
238 & $-2\cdot10^{-3}$ & 0.02 & 206 & $5\cdot10^{-3}$ & 0.07 \\
            & 2 & 223 & $-1\cdot10^{-3}$ & 0.04 & 211 &$-1\cdot10^{-3}$   & 0.02 &
235 & $1\cdot10^{-3}$  & 0.04 & 205 & $6\cdot10^{-3}$ & 0.08 \\
            & 3 & 225 & $2\cdot10^{-3}$  & 0.04 & 214 &  $-2\cdot10^{-3}$ & 0.04 &
231 & $-2\cdot10^{-3}$ & 0.04 & 208 & $5\cdot10^{-3}$ & 0.08 \\
            & 4 & 212 & $8\cdot10^{-3}$  & 0.08 & 206 & $3\cdot10^{-3}$   & 0.08 &
220 & $4\cdot10^{-3}$  & 0.08 & 180 & $2\cdot10^{-2}$ & 0.11 \\
\noalign{\smallskip}
\hline
\noalign{\bigskip}
\end{tabular}
\begin{minipage}{17cm}
  NOTE. Col.(1): Photometric parameter;
  Col.(2): Run of the Monte Carlo simulation.  Artificial galaxies are
  analyzed by assuming the correct values of PSF {\em FWHM} and sky
  level (Run 1), correct sky level and a PSF {\em FWHM} $2\%$ larger
  with respect to the actual one (Run 2), correct sky level and a PSF
  {\em FWHM} $2\%$ smaller with respect to the actual one (Run
  3). Artificial galaxies are built by assuming the correct PSF {\em
  FWHM} and a sky level $1\%$ larger with respect to the actual one
  (Run 4);
  Cols.(3, 6, 9, 12): Number of artificial galaxies in the magnitude bin;
  Cols.(4, 7, 10, 13): Mean of the relative errors;
  Cols.(5, 8, 11, 14): Standard deviation of the relative errors.
\end{minipage}
\end{center}
\end{scriptsize}
\end{table*}
 
\subsection{Results and comparison with previous studies} 
\label{sec:results_photometry} 
 
The parameters derived for the structural components of the sample
galaxies are collected in Table \ref{tab:sample}. All the listed
values are corrected for seeing smearing and galaxy
inclination. Surface brightnesses were calibrated by adopting, for the
2MASS images, the flux zero point given in the image headers
\citep{Jarrett00}.
 
The comparison of the structural parameters obtained for the same 
galaxy by different authors is often not straightforward on account of 
possible differences in the observed bandpass, parameterization of the 
surface-brightness distribution, and fitting method. 
 
\begin{figure} 
\centering 
\includegraphics[width=7cm,angle=90]{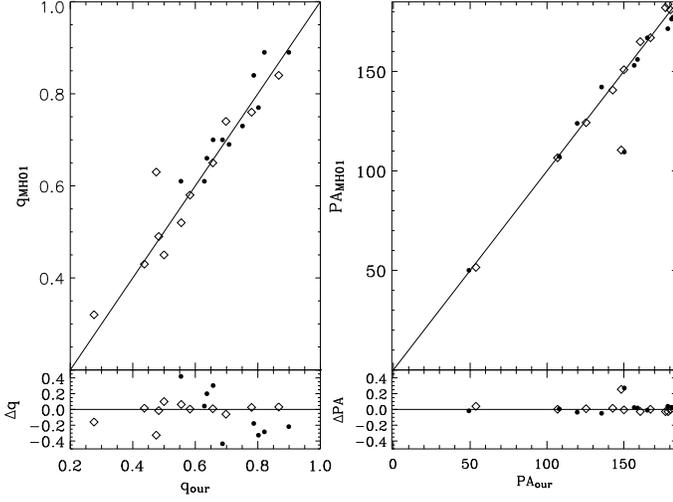} 
\caption{Comparison between the axis ratios (left panel) 
  and the position angles (right panel) measured in this paper 
  and by MH01. Filled dots and open diamonds correspond to 
  values measured for bulges and disks, respectively.  Residuals 
  $\Delta q$ and $\rm \Delta PA$ are defined as $1 - q_{\rm 
  \, MH01}/q_{\rm our}$ and $\rm 1 - PA_{\rm MH01}/PA_{\rm our}$, 
  respectively.} 
\label{fig:comparison_MH01} 
\end{figure}

MH01 already studied 11 of our sample galaxies (NGC~772, NGC~2775,
NGC~2841, NGC~2985, NGC~3169, NGC~3626, NGC~3675, NGC~3898, NGC~4450,
NGC~4501 and NGC~4826). They performed a two-dimensional parametric
decomposition of the $J-$band surface brightness distribution. They
considered ellipticities and position angles of both the bulge and
disk as free parameters. Therefore, we considered their results as the
most suitable to be compared with ours. The structural parameters we
measured are consistent with those given by MH01, within $25\%$ for
all the common galaxies but NGC~4826. We argue that they strongly
underestimated the scale length of its disk (and consequently obtained
a wrong estimate of the other parameters) because of the small field
of view ($3'\times3'$) of their image.
In Fig. \ref{fig:comparison_MH01} we show the comparison between our
axial ratios and position angles of the bulge and disk and those
measured by MH01.

\section{Correlations between structural parameters} 
\label{sec:correlations} 
 
The study of correlations between the structural parameters of bulges 
and disks of our sample galaxies will help us to both cross check our 
results with those available in literature and identify and rule out 
peculiar bulges from any further analysis. 

\subsection{Bulge parameters} 
\label{sec:bulges} 
 
\begin{figure*} 
\centering 
\includegraphics{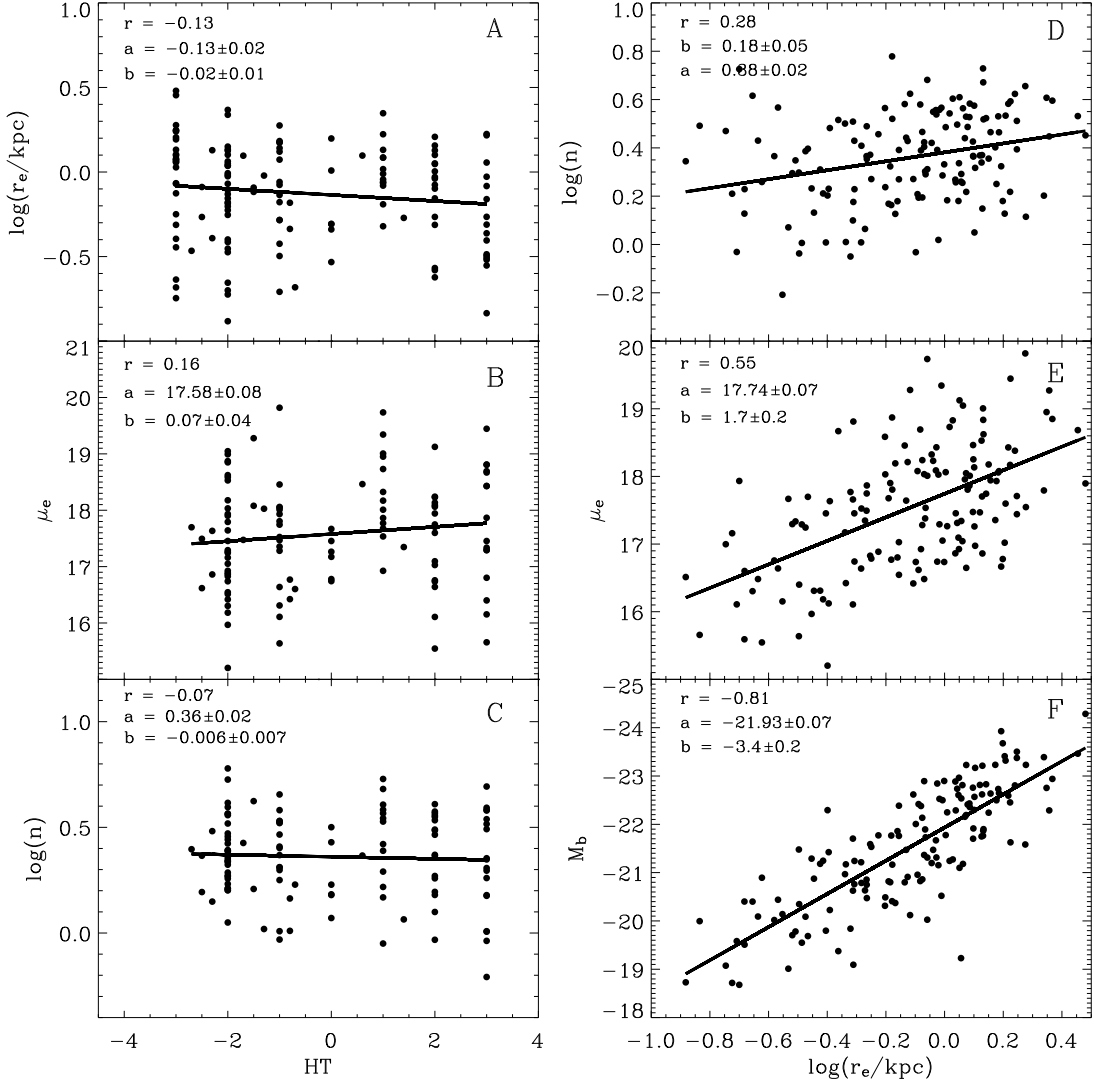} 
      \caption{Correlations between the bulge parameters. Correlation
      between the Hubble type and effective radius (A), effective
      surface brightness (B), and shape parameter (C). Correlations
      between the effective radius and shape parameter (D), effective
      surface brightness (E), and absolute magnitude (F). In each
      panel the solid line represent the linear regression through all
      the points. The Pearson correlation coefficient ($r$), and the
      results of the linear fit ($y =a + bx$) are also given.
      \label{fig:bulge_correlations}} 
\end{figure*}

We did not find any correlations between the bulge parameters and Hubble 
type. Neither the effective radius 
(Fig. \ref{fig:bulge_correlations}A), effective surface brightness 
(Fig. \ref{fig:bulge_correlations}B) nor the $n$ shape parameter 
(Fig. \ref{fig:bulge_correlations}C) show a statistically significant 
Pearson correlation coefficient ($r$). 
 
From near-infrared observations of spiral galaxies, Andredakis et
al. (1995) found that bulges of early-type spirals are characterized
by $n\approx4$ (i.e., they have a de Vaucouleurs radial surface
brightness profile), while the bulges of late-type spirals are
characterized by $n\approx1$ (i.e., they have an exponential radial
surface brightness profile).  This early result was confirmed in
various studies (e.g., de Jong et al. 1996; Khosroshahi et al. 2000;
MacArthur et al. 2003; M\"ollenhoff \& Heidt 2001; M\"ollenhoff 2004;
Hunt et al. 2004). We argue that our data does not show such a
correlation due to the smaller range of Hubble types covered by our
sample (S0--Sb) with respect to the cited works, where it is mostly
evident for Hubble types later than Sb.
 
The $n$ shape parameter increases with effective radius. Larger bulges
have a surface-brightness radial profile that which is more centrally
peaked than that of the smaller bulges
(Fig. \ref{fig:bulge_correlations}D). We obtained
 
\begin{equation} 
\log n = 0.38 (\pm 0.02) + 0.18 (\pm 0.05)\,\log r_{\rm e}\ \qquad (r = 0.28).  
\end{equation} 
%
 
The effective surface brightness is dependent on the effective radius. 
Larger bulges have a lower effective surface brightness 
(Fig. \ref{fig:bulge_correlations}E). We found a linear regression 
%
\begin{equation} 
\log \mu_{\rm e} = 17.74 (\pm 0.07) + 1.7 (\pm 0.2)\,\log r_{\rm e}\ \qquad  (r = 0.55).  
\end{equation} 
%
 
This is in agreement, within the errors, with the correlation by
MH01. If we use the mean surface brightness inside one effective
radius instead of the effective surface brightness this relation
becomes the so-called Kormendy relation, already known for bulges and
elliptical galaxies \citep{kormendy77}.
 
Finally, the absolute luminosity of the bulge is correlated with the
effective radius. Larger bulges are more luminous
(Fig. \ref{fig:bulge_correlations}F). This result in
 
\begin{equation} 
M_{\rm b} = -21.93 (\pm 0.06) - 3.4 (\pm 0.2)\,\log r_{\rm e}\ \qquad  (r = -0.80), 
\end{equation} 
%
where $M_{\rm b}$ is the $J-$band magnitude of the bulge.
A similar result was obtained by MH01 for a smaller sample of disk 
galaxies spanning a larger range of Hubble types. 
 
\begin{figure} 
\centering 
\includegraphics[width=8.5cm]{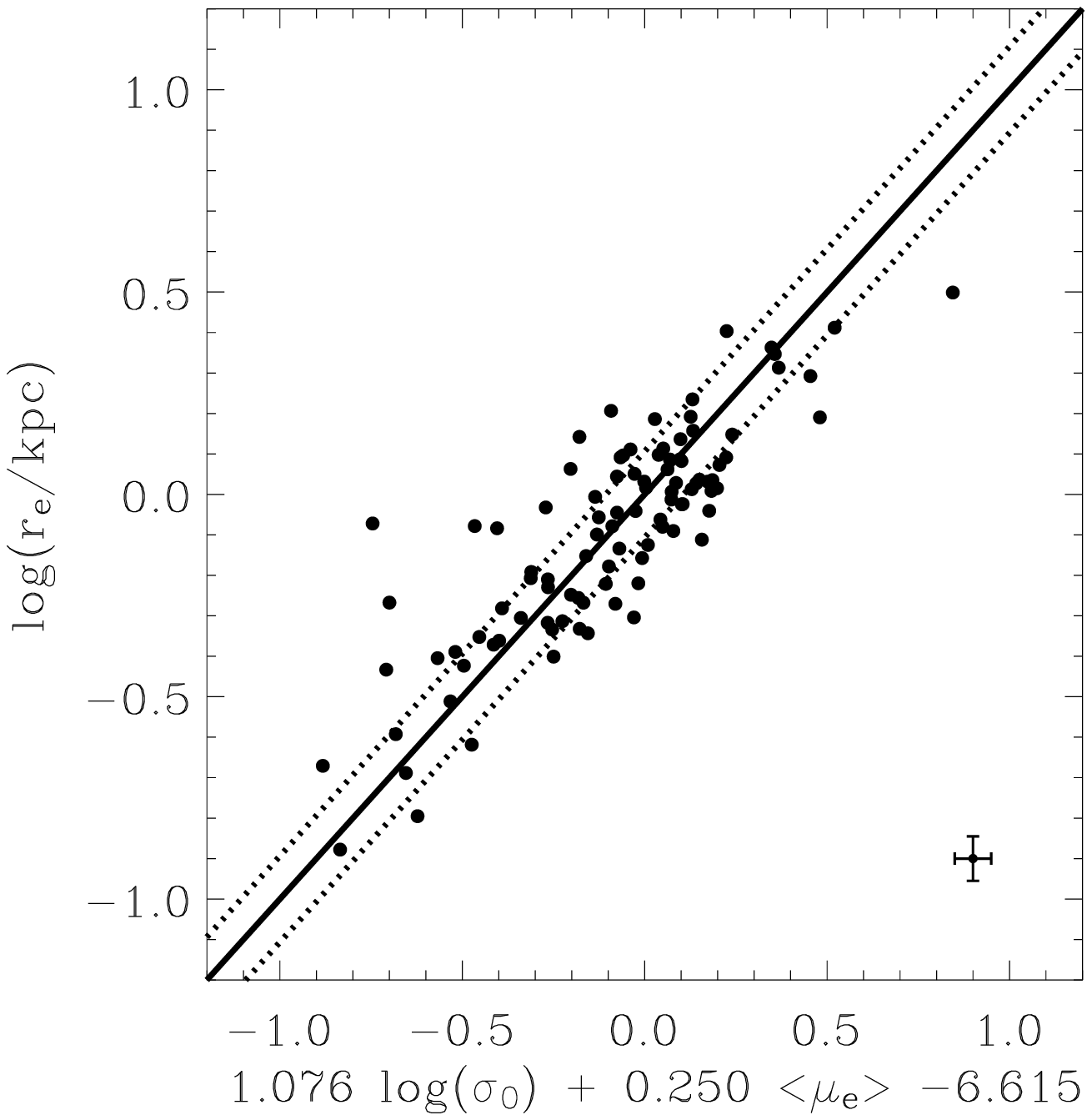} 
\caption{Edge-on view of the FP for the 98 early-to-intermediate type
  bulges of our sample with measured velocity dispersion. The solid
  line represents the linear fit to the data.  The dotted lines
  represent the $1\sigma$ deviation from the fit. The error bars in
  the lower right corner of the panel indicate the mean errors of the
  data.}
\label{fig:FP} 
\end{figure}

Bulges and elliptical galaxies follow a tight relation, the FP,
defined by the effective radius, mean surface brightness within
effective radius, and central velocity dispersion
\citep{djorgovskidavies87, dressler87}. Therefore, we derived the FP
for the bulges of our sample galaxies.
The measurements of the central stellar velocity dispersion for a 
subsample of 98 galaxies were available in literature and were 
retrieved from the on line HyperLeda catalog \citep{paturel03}.  
The velocity dispersions given by the catalogue are corrected to a
circular aperture of radius of 0.595 h$^{-1}$ kpc, which is equivalent
to an angular diameter of $3\farcs4$ at the distance of Coma,
following the prescription by Jorgensen et al. (1995).
The aperture-corrected velocity dispersions are given in Table 
\ref{tab:sample}. 
The coefficients describing the FP 
%
\begin{equation} 
\log r_{\rm e}=1.08 (\pm 0.09) \log \sigma_0 + 0.25 (\pm 0.02)  
  \langle \mu_{\rm e} \rangle - 6.61 (\pm 0.40),
\end{equation} 
%
were derived by minimizing the square root of the residuals along the
$\log r_{\rm e}$ axis. Errors given for every coefficient were
calculated by performing a bootstrap analysis with 1000 iterations. No
statistically significant difference was observed when only bulges of
lenticular or early-to-intermediate spiral galaxies were
considered. The dispersion around this relation is $\sigma=0.11$ dex
and was measured as the rms scatter in the residuals of $\log r_{\rm
e}$.
The observational error on the FP is 0.066 dex and includes the
measurement errors in $\log r_{\rm e}$ (0.055 dex), $\log \sigma_0$
(0.029 dex), and $\langle \mu_{\rm e} \rangle$ (0.021 mag
arcsec$^{-2}$). Errors in $\log r_{\rm e}$ and $\langle \mu_{\rm e}
\rangle$ are not independent (Kormendy 1977). Compared with the
dispersion around the relation, this gives an intrinsic scatter of
0.088 dex.
Figure \ref{fig:FP} shows an edge-on view of the FP. Our coefficients
and those by Falc\'on-Barroso et al. (2002) are consistent within the
errors, although they analyzed $K-$band data. Unfortunately, we have
not found in the literature the coefficient of the FP in the $J-$band
for a direct comparison \citep[see][]{bernardi03}.
 
One of the projections of the FP is the so-called Faber-Jackson
relation (FJ), which relates the luminosity of elliptical galaxies and
bulges to their central velocity dispersion \citep{faberjackson76}.
We derived the $J-$band FJ relation for the bulge subsample obtaining 
%
\begin{equation} 
\log \sigma_0 = 0.1 (\pm 0.2) - 0.095 (\pm 0.009) M_{\rm b}\ \qquad  (r = -0.71). 
\label{eq:FJ_all}
\end{equation} 
%
 
This result also holds when we consider only galaxies with errors on
the central velocity dispersion lower than 10 km~s$^{-1}$ ($\Delta
\log \sigma_0<0.018$ dex). In fact, we derived
%
\begin{equation} 
\log \sigma_0 = 0.4 (\pm 0.3) - 0.085 (\pm 0.013) M_{\rm b}\ \qquad (r = -0.65),
\end{equation} 
%
which is consistent within errors with Eq. \ref{eq:FJ_all}. From
Eq. \ref{eq:FJ_all} we derived $L \propto \sigma_0^{4.2}$, which is
very close to the virial relation and indicates that our bulges share
important characteristics with bright elliptical galaxies
\cite{matkovicguzman05}. On the other hand, Balcells et al. (2007)
found $L \propto \sigma_0^{2.9}$ (close to faint ellipticals)
observing a sample of bulges with the Hubble Space Telescope in the
$K-$band.
This discrepancy is not due to the adopted fitting method and it
is not observed if only bright bulges ($M_{\rm b}<-20$ mag) are
considered. Indeed, we found $L \propto \sigma_0^{(3.9\pm0.4)}$ for
our sample and $L \propto \sigma_0^{(3.6\pm0.7)}$ for the sample by
Balcells et al. (2007). The different behaviour of faint bulges
($M_{\rm b}>-20$ mag) requires further investigation to be explained.

\begin{figure} 
\centering 
\includegraphics[width=8.5cm]{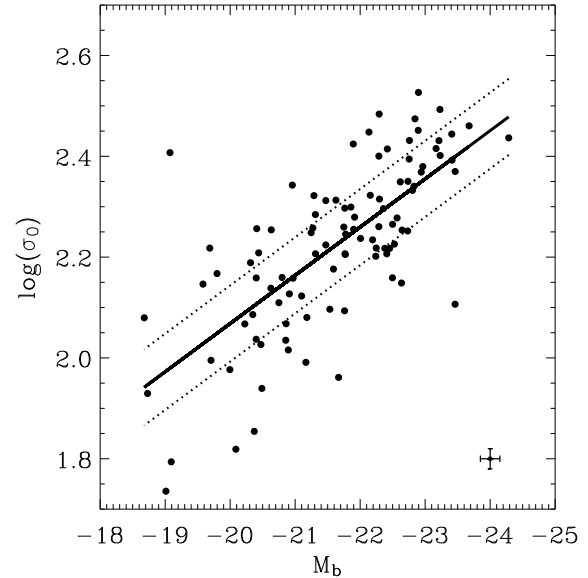} 
\caption{FJ relation for the 98 early-to-intermediate type bulges of
  our sample with measured velocity dispersion. The solid line
  represents the linear fit to the data. The dotted lines represent
  the $1\sigma$ deviation from the fit. The error bars in the lower
  right corner of the panel indicate the mean errors of the data.}
\label{fig:FJ} 
\end{figure} 
 
\begin{figure} 
\centering 
\includegraphics[width=8.5cm]{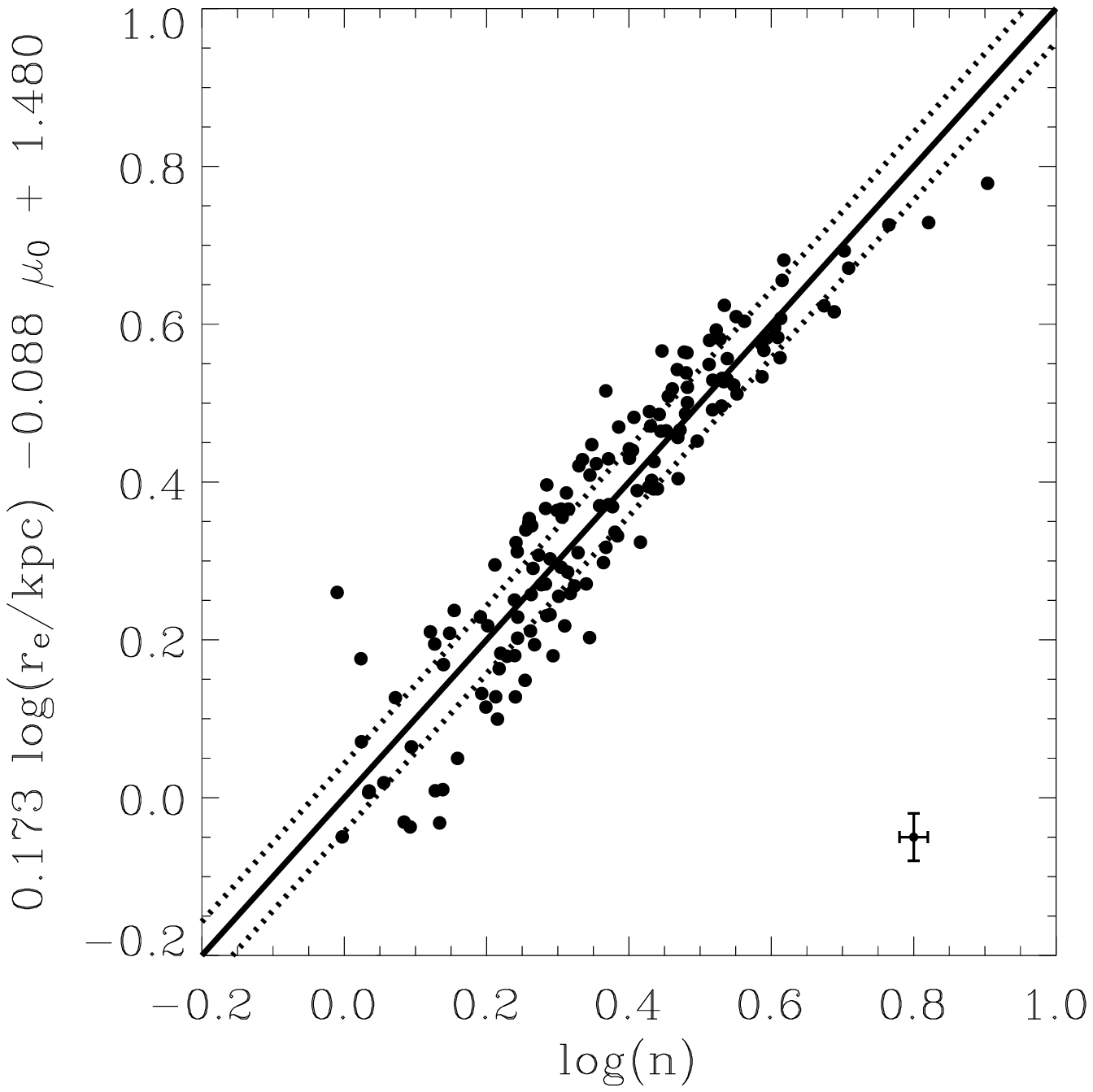} 
\caption{Edge-on view of the PP for early-to-intermediate bulges of
  our sample. The solid line represents the linear fit to the data.
  The dotted lines represent the $1\sigma$ deviation from the fit. The
  error bars in the lower right corner of the panel indicate the mean
  errors of the data.}
\label{fig:PP} 
\end{figure}

Khosroshahi et al. (2000) noticed that the shape parameter, effective
radius and central surface brightness of elliptical galaxies and
bulges are correlated. This relation was termed photometric plane
(PP). Figure \ref{fig:PP} shows an edge-on view of the PP of our bulge
sample
%
\begin{equation} 
\log n = 0.17 (\pm 0.02) \log r_{\rm e} - 0.088 (\pm 0.004) \mu_0  
+ 1.48 (\pm 0.05). 
\end{equation} 
%

The coefficients were derived by minimizing the square root of the
residuals along the $\log n$ axis. Errors given for every coefficient
were calculated by performing a bootstrap analysis with 1000
iterations. No statistically significant difference was observed when
only bulges of lenticular or early-to-intermediate spiral galaxies
were considered. The dispersion around this relation is $\sigma =
0.04$ and was measured as the rms scatter in the residuals of $\log
n$.  Our coefficients and those by MH01 are consistent within the
errors, although their sample is dominated by bulges of late-type
spirals. The presence of the bulges of lenticular and spirals on the
same PP hints towards a common formation scenario.
 
\subsection{Disk parameters} 
\label{sec:disks} 
 
\begin{figure*} 
\centering 
\includegraphics{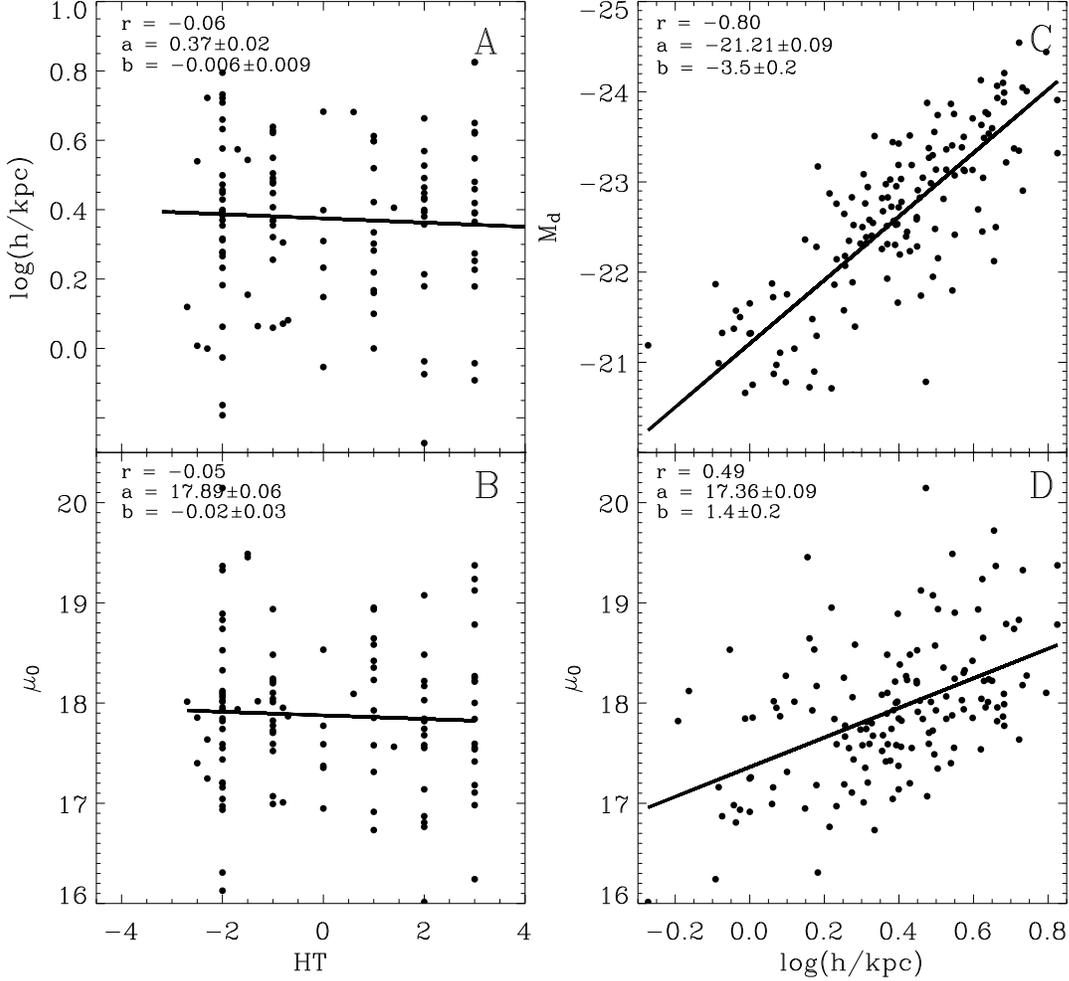} 
      \caption{Correlations between the disk parameters. Correlations
      between the Hubble type and disk scale length (A) and central
      surface brightness (B).  Correlation between the disk scale
      length and absolute luminosity (C) and central surface
      brightness (D). Solid lines and coefficients as in
      Fig. \ref{fig:bulge_correlations}.}
\label{fig:disk_correlations} 
\end{figure*}

Regarding the disk parameters, we found no correlation between the
scale length and Hubble type ($r = -0.06$,
Fig. \ref{fig:disk_correlations}A). The same is true for the central
surface brightness. In fact, it shows a large scatter also with Hubble
type ($r = -0.05 $, Fig. \ref{fig:disk_correlations}B). This is
consistent with the results of de Jong et al. (1996) and MH01.
 
On the other hand, the central surface brightness and the luminosity
of the disks are dependent on the scale length. Larger disks have a
lower central surface brightness
(Fig. \ref{fig:disk_correlations}D). We found a linear regression
%
\begin{equation} 
\log \mu_{\rm 0} = 17.36 (\pm 0.1) + 1.4 (\pm 0.2)\,\log h\ \qquad (r = 0.49), 
\end{equation} 
%
and brighter disks show larger scale lengths 
(Fig. \ref{fig:disk_correlations}C)  
 
\begin{equation} 
M_{\rm d} = -21.21 (\pm 0.09) - 3.5 (\pm 0.2)\,\log h\ \qquad (r =
-0.80),
\end{equation} 
%
where $M_{\rm d}$ is the $J-$band magnitude of the disk.

The coefficients are in agreement 
within the errors with those given by MH01. 
 
\subsection{Bulge and disk interplay} 
\label{sec:bulgesdisks} 
 
\begin{figure*} 
\centering 
\includegraphics{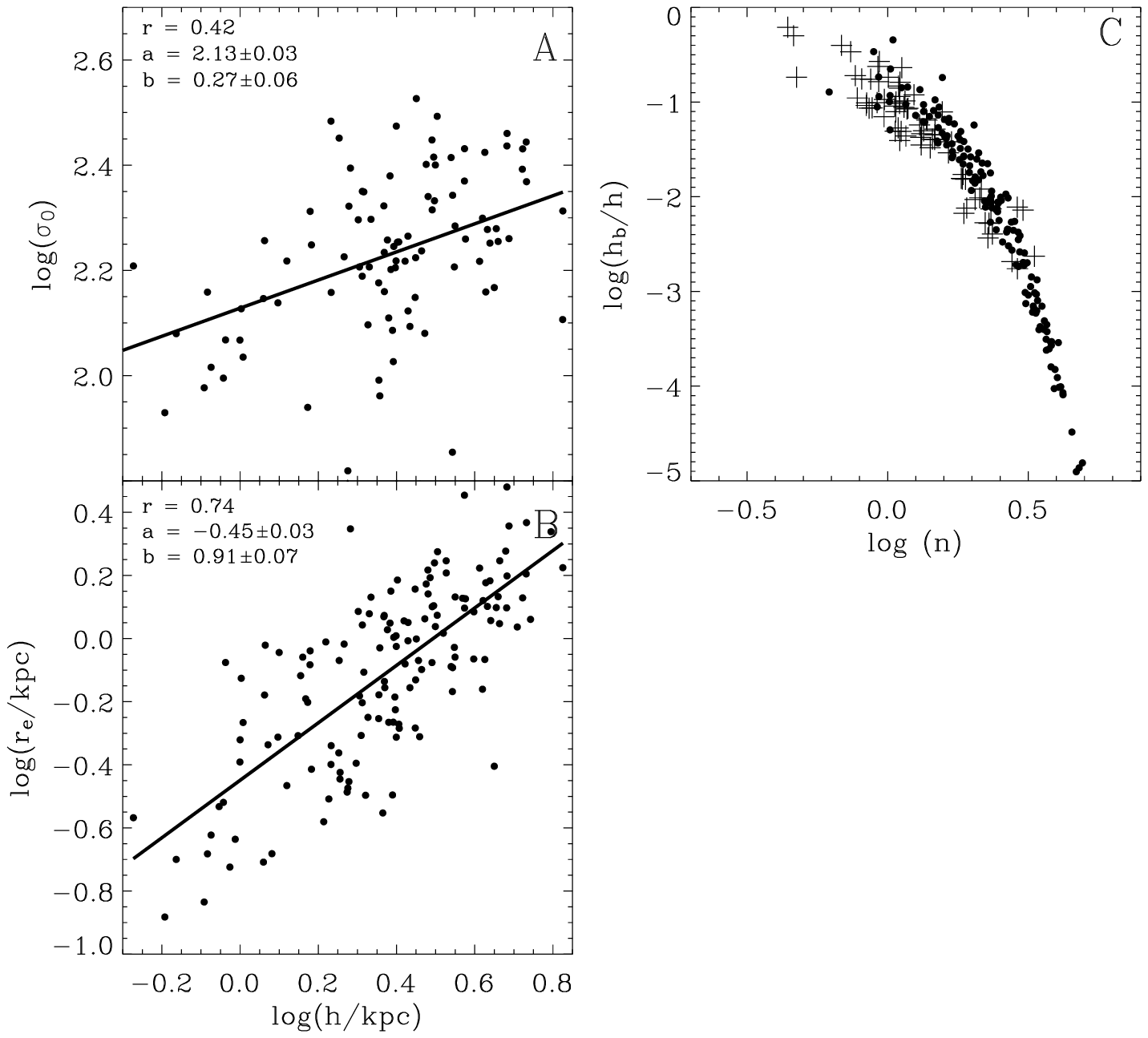} 
      \caption{(A) Correlation between the disk scale length and
      central velocity dispersion for the 98 galaxies of our sample
      with a measured velocity dispersion. (B) Correlation between the
      disk scale length and bulge effective radius. (C) The ratio
      between the bulge and disk exponential scale lengths as a
      function of the bulge shape parameter.  Filled circles and
      crosses represent the results of our measurements and
      simulations by Tissera et al. (2006), respectively. Solid lines
      and coefficients as in Fig. \ref{fig:bulge_correlations}}
      \label{fig:bulgedisk_interplay} 
\end{figure*} 
 
We have found that the disk scale length increases with central 
velocity dispersion. Since central velocity dispersion correlates with the
virial mass of the bulge (${\cal M}_{\rm b} = \alpha r_{\rm e} \sigma_0^2/G$ 
with $\alpha =5$, Cappellari et al. 2006), we conclude that larger disks 
are located in galaxies with more massive bulges 
(Fig. \ref{fig:bulgedisk_interplay}A). For the subsample of 98 
early-to-intermediate bulges with a measured velocity dispersion we 
found 
%
\begin{equation} 
\log \sigma_0 = 2.13 (\pm 0.03) + 0.27 (\pm 0.06)\,\log h\ \qquad (r = 0.42). 
\end{equation} 
%
 
We also found a strong correlation between the bulge effective radius 
and the disk scale length.  Larger bulges reside in larger disks 
(Fig. \ref{fig:bulgedisk_interplay}B).  This relation was already 
observed by Courteau et al. (1996) and later observed in NIR by MH01, 
MacArthur et al. (2003).  We obtained a linear regression 
%
\begin{equation} 
\log r_{\rm e} = -0.45 (\pm 0.03) + 0.91 (\pm 0.07)\,\log h\ \qquad (r = 0.74), 
\end{equation} 
%
which is in agreement within error bars with the correlation found by 
MH01. 
All these correlations between bulge and disk parameters indicate a
link between the bulge and disk formation and evolution history. This
connection was interpreted as an indication of the formation of
late-type bulges via secular evolution of the disks
\citep{courteau96}. However, our measurements of the scale lengths of
the bulge and disk and $n$ (Fig. \ref{fig:bulgedisk_interplay}C) are
also fully consistent with the predictions of the numerical
simulations by Scannapieco \& Tissera (2003) and Tissera et
al. (2006). They studied the effects of mergers on the mass
distribution of bulges and disks of galaxies formed in hierarchical
clustering scenarios. Our mean value $\langle r_{\rm e}/h\rangle=0.36
\pm 0.17$ is also in good agreement with $r_{\rm e}/h$ found by Naab
\& Trujillo (2006) for a series of major mergers' remnants. These
results indicate that these relations are not enough to distinguish
between bulges formed by mergers or by secular evolution of the disk,
even if a strong crosstalk between both components is present.
 
\section{The equatorial intrinsic ellipticity of bulges} 
\label{sec:ellipticity} 
 
In the previous section, we realized that the knowledge of
correlations between structural parameters of the bulge and disk is
not sufficient to distinguish between the different scenarios that
were proposed to explain the formation of bulges. Therefore, we
decided to study the intrinsic shape of bulges in order to give a
further constraint to these scenarios.
 
Independently of its internal structure, we can consider a bulge of a
spiral galaxy as an ellipsoidal stellar system located in the center
of the galaxy,which stands out against the disk in the photometric
observations. We assume that both the bulge and disk share the same
center, which coincides with the galactic one, and they have the same
polar axis (i.e., the equatorial plane of disk coincides with that of
bulge).
 
\subsection{Geometrical formalism} 
\label{sec:formalism} 
 
Let $(x,y,z)$ be Cartesian coordinates with the origin in the galaxy
center, the $x-$axis and $y-$axis corresponding to the principal axes
of the bulge equatorial ellipse, and the $z-$axis corresponding to the
polar axis. As the equatorial plane of the bulge coincides with the
equatorial plane of the disk, the $z-$axis is also the polar axis of
the disk.  If $A$, $B$, and $C$ are lengths of the ellipsoid
semi-axes, the corresponding equation of the bulge in its own
reference system is given by
 
\begin{equation} 
\frac{x^2}{A^2} + \frac{y^2}{B^2} + \frac{z^2}{C^2} = 1. 
\label{eqn:ellipsoid} 
\end{equation} 
 
Let $(x',y',z')$ be now the Cartesian coordinates of the observer
system. It has its origin in the galaxy center, the polar $z'-$axis
along the line-of-sight (LOS) and pointing toward the galaxy, and the
plane of the sky lies on the $(x',y')$ plane.
 
The projection of the disk onto the sky plane is an ellipse whose
major axis is the line of nodes (LON), i.e., the intersection between
the galactic and the sky planes. The angle $\theta$ between the
$z-$axis and $z'-$axis corresponds to the inclination of the bulge
ellipsoid; it can be derived as $\theta=\arccos{(d/c)}$ from the
length $c$ and $d$ of the two semi-axes of the projected ellipse of
the disk.
We defined $\phi$ ($0 \leq \phi \leq \pi /2$) as the angle between the 
$x$-axis and the LON on the equatorial plane of the bulge 
$(x,y)$. 
Finally, we also defined $\psi$ ($0 \leq \psi \leq \pi /2$) as the
angle between the $x'$-axis and the LON on the sky plane
$(x',y')$. The three angles $\theta$, $\phi$, and $\psi$ are the usual
Euler angles and relate the reference system $(x,y,z)$ of the
ellipsoid with that $(x',y',z')$ of the observer by means of three
rotations. Indeed, since the location of the LON is known, we can
choose the $x'-$axis along it, and consequently it holds that
$\psi=0$.
By applying these two rotations to Eq. \ref{eqn:ellipsoid} it is 
possible to derive the equation of the ellipsoidal bulge in the 
reference system of the observer, as well as the equation of the 
ellipse corresponding to its projection on the sky plane 
\citep{simonneau98}. 
Now, if we identify this ellipse with the ellipse that forms the
observed ellipsoidal bulge, we can determine the corresponding axes of
symmetry x$_e$ and y$_e$. The first one, on which we measured the
semi-axis $a$, forms an angle $\delta$ with the LON (the x'-axis in
the observed plane); the semi-axis $b$ is taken over the
y$_e$-axis. We always choose $0 \le \delta \le \pi/2$, so it is
possible that $a$ either be the major or the minor semi-axis, and vice
versa for $b$.
 
We have 
 
\begin{scriptsize} 
\begin{eqnarray} 
a^{2}b^{2} &=& A^{2}C^{2}\sin^{2}\theta \cos^{2}\phi + 
  B^{2}C^{2}\sin^{2}\theta \sin^{2}\phi + A^{2}B^{2}\cos^{2}\theta. 
\label{eqn:ab}\\ 
a^{2}+b^{2}&=&A^{2}(\cos^{2}\phi + \cos^{2}\theta \sin^{2}\phi) +  
  B^{2}(\sin^{2}\phi + \cos^{2}\theta \cos^{2}\phi) + C^{2}\sin^{2}\theta.  
\label{eqn:a+b}\\ 
\tan2\delta&=&\frac{(B^{2}-A^{2})\cos\theta\sin2\phi}{A^{2}(\cos^{2} 
  \theta\sin^{2}\phi -\cos^{2}\phi) +  
  B^{2}(\cos^{2}\theta\cos^{2}\phi - \sin^{2}) + C^{2}\sin^{2}\theta}. 
\label{eqn:tandelta} 
\end{eqnarray} 
\end{scriptsize} 
 
If the ellipsoidal bulge is triaxial ($A \ne B \ne C$) then it is
possible to observe a twist ($\delta\ne0$; see Eq. \ref{eqn:tandelta})
between the axes of the projected ellipses of the bulge and the disk.
 
\subsection{Inverse problem or deprojection} 
\label{sec:deprojection} 
 
We will now focus our attention on the inverse problem, i.e.,
deprojection. Following Simonneau et al. (1998), from
Eqs. \ref{eqn:ab}, \ref{eqn:a+b}, and \ref{eqn:tandelta}, we are able
to express the length of the bulge semi-axes (i.e. $A$, $B$, and $C$)
as a function of the length of the semi-axes (i.e. $a$, $b$) of the
projected ellipse and the position angle ($\delta$).
 
\begin{small} 
\begin{eqnarray} 
A^2 & = & \frac{a^2+b^2}{2}\left[ 1 + e\left( cos2\delta + sin2\delta\frac{sin\phi}{cos\phi}\frac{1}{cos\theta}\right)\right]  \label{eqn:A} \\ 
B^2 & = & \frac{a^2+b^2}{2}\left[ 1 + e\left( cos2\delta - sin2\delta\frac{cos \phi}{sin \phi}\frac{1}{cos\theta}\right)\right] \label{eqn:B} \\ 
C^2 & = & \frac{a^2+b^2}{2}\left[ 1 + e\left( 2sin2\delta \frac{cos\theta}{sin^2\theta} \frac{cos2\phi}{sin2\phi} - cos2\delta\frac{1 + cos^2\theta}{sin^2\theta} \right)\right], \label{eqn:C}
\end{eqnarray} 
\end{small} 
%
where $e=(a^2 - b^2)/(a^2 + b^2)$ is, in some way, a measure of the 
ellipticity. It will be $-1 \le e \le 1$. 
 
Notice that $a,$ $b,$ $\delta$ and $\theta$ are all observed
variables. Unfortunately, the relation between the intrinsic and
projected variables also depends on the spatial position of the bulge
(i.e., on the $\phi$ angle), which is not directly accessible to
observations. For this reason, only a statistical determination can be
performed to assess the intrinsic shape of bulges.
 
As $A$ and $B$ are the semi-axis of the equatorial ellipse of the
bulge, we have to distinguish between two cases, according to
Eqs. \ref{eqn:A} and \ref{eqn:B}. If $a>b$ (or equivalently $e>0$)
then $A>B$. Otherwise, if $a<b$ (or equivalently $e<0$) then $A<B$.
Thus, if $\delta \neq 0$ the equatorial plane of the bulge ellipsoid
is not circular and the bulge ellipsoid is triaxial.  
 
From Eqs. \ref{eqn:A}, \ref{eqn:B} and \ref{eqn:C} we can write the
axial ratios $A/C$ and $B/C$ as explicit functions of
$\phi$. Moreover, we assume that the angle $\phi$ is random and
independent of the length of ellipsoid semi-axes. Thus, the normalized
probability distribution $P(\phi)$ of getting a given value of $\phi$ in
$(\phi,\phi$+$d\phi)$ is
 
\begin{equation} 
P(\phi)=2/\pi\ \ ; \qquad \int_{0}^{\pi/2} P(\phi) d\phi = 1.  
\label{eqn:phi_prob} 
\end{equation} 
 
According to a fundamental theorem of statistics, the probability of
obtaining a given value of any function $f(\phi)$ (e.g., one of the
axial ratios) will be equal to the probability of getting the
corresponding value of $\phi$, provided that the ratio $df/d\phi$
between the corresponding differential elements is taken into account.
 
In this work, we will focus our attention on the intrinsic equatorial
ellipticity of bulges. We define it as
$E=($$A^{2}$-$B^{2}$)/($A^{2}$+$B^{2})$ with $-1$$<$$E$$<$ $1$. In a
forthcoming paper, we will study the intrinsic flattening of the bulge
ellipsoids defined as the ratio between the length $C$ of polar
semi-axis and the mean length of the equatorial semi-axes.
 
From Eqs. \ref{eqn:A} and \ref{eqn:B}, it is straightforward to derive 
a relation among the intrinsic variables (equatorial ellipticity $E$ 
and position angle $\phi$), and the measured ( i.e. $\theta$, $e$, and 
$\delta$), which is 
 
\begin{equation} 
\frac{E\, \sin(2\phi)}{1 + E\, \cos(2\phi)} =  
\frac{1}{\cos\theta}\frac{e\sin2\delta}{1 + e\cos2\delta} \equiv Q. 
\label{eqn:qe} 
\end{equation} 
%

The second member of the equality in Eq. \ref{eqn:qe} allows us to define 
the observable $Q$ in terms of the measured variables $\theta$, $e$, 
and $\delta$. It must be stressed that for each specific bulge the 
relation between the equatorial ellipticity $E$ and the unknown 
parameter $\phi$ embraces the whole of the measured variables through 
the single variable $Q$. 
 
On the one hand, Eq. \ref{eqn:qe} will yield the conditional
probability $P_{Q}(E)$ that a given bulge, with a measured value of
$Q$, takes on any particular value of $E$ (individual statistic); on
the other hand, this equation will give the probability $P_{E}(Q)$
associated to each value of $Q$ for a bulge with intrinsic equatorial
ellipticity $E$. This latter probability will be the kernel of an
integral equation that relates the observed statistical distribution
$P(Q)$, corresponding to a sample of galaxies, with the statistical
distribution of the equatorial ellipticity $P(E)$ for the same sample.
 
\subsection{Individual statistics}  
\label{sec:individual} 
 
For a given galaxy, we can measure the values of $\theta$, $e$ and
$\delta$, and then derive the value of $Q$ through
Eq. \ref{eqn:qe}. We want to determine the probability $P_{Q}(E)$ that
such a galaxy (i.e. with such a value of $Q$) will take on a value of $E$
in the range $(E, E$+$dE)$.  The subindex $Q$ specifies this
galaxy. All the galaxies with the same value of $Q$ shall partake the
same probability distribution $P_{Q}(E)$.
 
Once the value of $Q$ is prescribed, for some values of $E$ there are
not values of $\phi$ ($0 \le \phi \le \pi/2$) that satisfy
Eq. \ref{eqn:qe}. Hence, it shall hold that $P_{Q}(E)=0$. Only for
those values of $E$ such that
 
\begin{equation} 
E^{2} \ge \frac{Q^2}{1 + Q^2} \equiv T^{2}\ \ ;\ \  \qquad  0 \le T \le 1,
\label{eqn:Tdef} 
\end{equation} 
will there will exist two values of $\phi$ that fulfill
Eq. \ref{eqn:qe}.
 
Then for any value of $E > T$ the probability $P_{Q}(E)$ will be given by 
 
\begin{equation} 
P_{Q}(E)=\sum_{j=1,2} \left(P(\phi) \left\arrowvert  
\frac{\delta \phi}{\delta E} \right\arrowvert_{Q}\right)_{\phi_{j}}\ . 
\label{eqn:ind_estat} 
\end{equation} 
%

By calculating the partial derivative of Eq. \ref{eqn:ind_estat} and
normalizing we obtain
 
\begin{equation} 
P_{T}(E)= \frac{\frac{T}{E}\frac{1}{\sqrt{E^{2}-T^{2}}}}{\arccos T}\ ,
\label{eqn:ind_stat_final} 
\end{equation} 
where the subindex $T$ plays the same role as the subindex $Q$; both 
are related through Eq. \ref{eqn:Tdef}. 
 
This means that the possible values of $E$ are very concentrated and
slightly larger than $T$. To get an idea of how $P_{T}(E)$ is peaked
near the value of $T$, we calculated the value $E_{1/2}$ for which the
total probability that $E > E_{1/2}$ is equal to the probability that
$E < E_{1/2}$. For every bulge $E_{1/2}$ is a sort of mean value of
$E$, and is given by
 
\begin{equation} 
E_{1/2}=\sqrt{\frac{2}{1+T}}T. 
\label{eqn:e1/2} 
\end{equation} 
 
\begin{figure} 
\centering 
\includegraphics[width=8.5cm]{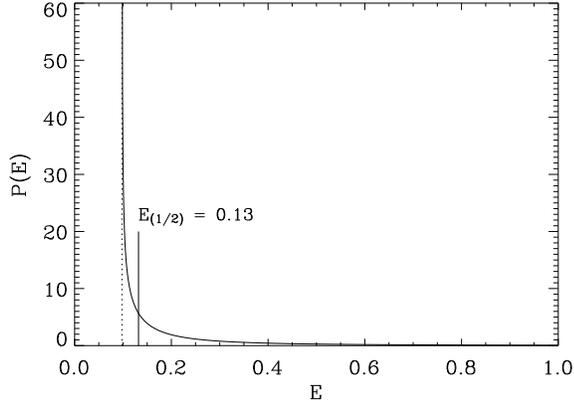} 
\caption{Probability distribution function of $E$ for the galaxy IC 
4310. The dotted line represents the value of T=0.098 derived for this 
galaxy. The value of $E_{1/2}$ is also shown in the plot.} 
\label{fig:PEindiv} 
\end{figure}

In Fig.\ref{fig:PEindiv} we show, as an example, the probability 
$P_T(E)$ for one galaxy of our sample. 
 
\subsection{Global statistics} 
\label{sec:global} 
 
Likewise, we can define the probability $P_E(Q)$ associated to each 
value of $Q$ for a given bulge of intrinsic equatorial ellipticity 
$E$. 
 
For a prescribed value of $E$, it will only be possible to get the
values of $\phi$ that satisfy Eq. \ref{eqn:qe} when $T \le E$. The
probability $P(\phi_{j})$ for the two values of $\phi_j$ is given by
Eq. \ref{eqn:phi_prob}. The probability $P_E(T)$ is equal to the sum
of the two probabilities $P(\phi_{j})$, weighted with the ratio
$(\delta \phi/\delta T)_{E}$ of the differential elements:
 
 \begin{equation} 
P_{E}(Q)=\sum_{j=1,2} \left(P(\phi) \left\arrowvert \frac{\delta 
\phi}{\delta Q} 
\right\arrowvert_{E}\right) _{\phi_{j}}. 
\label{eqn:ind_est} 
\end{equation} 
%
 
Once the partial derivative from Eq. \ref{eqn:qe} is computed, we
obtain
 
\begin{equation} 
P_{E}(T)= \frac{2}{\pi}\frac{1}{\sqrt{E^{2}-T^{2}}}\ \ ;\ \qquad E \ge T\ .  
\label{eqn:ind_stat_fin} 
\end{equation} 
 
Our purpose here is to determine the probability distribution $P(E)$,
that is the number of galaxies with a value of $E$ inside the interval
$(E, E+dE)$, starting from a sample that is sufficiently
representative. We have measured for such a sample the distribution
$P(T)$, namely the number of bulges whose values of $T$ are within $T$
and $T+dT$.  We must write now $P(T)$ as the integral over all the
values of $E$ of the product of the conditional probability $P_E(T)$
by the so far unknown distribution $P(E)$:
 
\begin{equation} 
P(T)=\int_{-1}^{1}P(E)\,P_{E}(T)\,dE. 
\label{eqn:probability} 
\end{equation} 
 
Then, by making use of Eq. \ref{eqn:ind_stat_fin}, we obtain an Abel-like 
integral equation 
 
\begin{equation} 
P(T)=\frac{2}{\pi}\int_{T}^{1}\frac{P(E)}{\sqrt{E^2 - T^2}}dE, 
\label{eqn:probability1} 
\end{equation} 
%
which will allow us to derive $P(E)$ from the observed distribution
$P(T)$.
 
However, as usual, the data $P(T)$ of our statistical problem takes
the form of histograms, hence, the relevant equations must be
formulated accordingly.
 
Let $T_{k}$ with $k=0, 1, 2, ..., N$ ($T_{0}=0$, $T_{N}=1$) be a set 
of discrete ordinates that defines the histogram of the observed 
function $P(T)$ (Fig. \ref{fig:PT}). The $k^{th}$-element of this histogram 
is defined by 
 
\begin{equation} 
P_{k}(T)=\frac{1}{T_{k}-T_{k-1}}\int_{T_{k-1}}^{T_{k}}P(T)\,dT. 
\label{eqn:probabili} 
\end{equation} 
%

We must now seek the integral equation that relates the variables 
$P_k(T)$ with the probability distribution $P(E)$. 
 
We notice that the integral of $P(T)$ in Eq. \ref{eqn:probabili} is
equivalent to the difference between the two quadratures of $P(T)$
over the intervals ($T_{k-1}$,1) and ($T_k$,1). In both of them, we
will replace $P(T)$ by its integral form, given by Eq.
\ref{eqn:probability1}.
 
Then, by inverting the order of integration, we can easily rewrite the 
two resulting integrals to obtain 
 
\[ 
P_{k}(T)=\frac{1}{T_k -T_{k-1}}\int_{T_{k-1}}^{T_k}\,P(E)\,dE\]  
\begin{equation} 
\ \ \ \ \ \ \ \ \ \ - \frac{2/\pi}{T_{k}-T_{k-1}}  
\left(M(T_{k-1}) - M(T_{k}) \right), 
\label{eqn:probabinueva} 
\end{equation} 
%
where we defined 
\begin{equation} 
M(T_{k})=\int_{T_{k}}^{1}P(E)\arcsin (T_k/E) \,dE,  
\label{eqn:MT}  
\end{equation} 
%
and $M(T_{k-1})$ in a similar way. 
 
Equation \ref{eqn:probabinueva} is the integral equation that will
allow us to obtain the values of $P(E)$. Since it is consistent with
the numerical structure of the data, we are confident that we have
eliminated most of the numerical problems that arise from the direct
inversion of Eq. \ref{eqn:probability1}, which constitutes a typical
ill-posed problem.
 
At this point, we require for $P(E)$ the same histogram representation
that we have already introduced for the data $P(T)$. We introduce a
similar set of discrete ordinates for the variable $E$ (i.e. $E_k \sim
T_k$), and an analogous definition to obtain the elements $P_k(E)$ of
the histogram
 
\begin{equation} 
P_{j}(E)=\frac{1}{E_{j}-E_{j-1}}\int_{E_{j-1}}^{E_{j}}P(E)\,dE. 
\label{eqn:probabiliE} 
\end{equation} 
%
 
Thus Eq. \ref{eqn:probabinueva} can be rewritten into the form 
 
\begin{equation} 
P_{k}(T)=P_{k}(E) - \frac{2/\pi}{T_{k}-T_{k-1}}  
\left(M(T_{k-1}) - M(T_{k}) \right). 
\label{eqn:probabi} 
\end{equation} 
%

In order to express the integrals $M(T_{k-1})$ and $M(T_{k})$ as
linear functions of the so far unknown values of $P_{j}(E)$ (where $j
> k-1$), we consider that $P(E)$ is constant and equal to the unknown
values of $P_{j}(E)$ over each interval ($E_{j-1}$, $E_j$), according
to its histogram representation. Therefore, it is straightforward to
compute the coefficients $CM (k, j)$, defined by the linear relation
 
\begin{equation} 
M(T_{k})=\sum_{j=k+1}^{N} CM(k,j)\, P_{j}(E). 
\label{eqn:cmdef} 
\end{equation} 
 
Thus Eq. \ref{eqn:probabi} becomes a simple linear algebraical
equation that relates the terms of the two histograms $P_k(T)$ and
$P_j(E)$ through a triangular matrix. Once we have the integral
Eq. \ref{eqn:probability1} into a suitable matrix form, according to
the histogram representation of the data and results, a simple matrix
inversion could be, in principle, enough to obtain the resulting
$P_j(E)$. However, such a procedure may add to the intrinsic
difficulties, due to the lack of precision typical of the
observational data, which naturally arise in the matrix inversion
process; a catastrophic mixture when dealing with an inverse problem.
 
\subsection{Inversion methods for the integral equation} 
\label{sec:methods} 
 
We have considered two different approaches to tackle the numerical
problem. The first one is suggested by the method that leads us to the
integral Eq. \ref{eqn:probabi} for the set of discrete elements
$P_{k}(T)$. We notice that there are two different terms in its
right-hand side. The first one is the identity operator. The second
one is the difference between two integrals of $P(T)$, multiplied by a
kernel that is quickly decreasing. We can consider the former as the
leading term, and treat the latter as a corrective term in a iterative
perturbation method.
Consequently, we can write Eq. \ref{eqn:probabi} as  
 
\begin{equation} 
P_{k}(E)=P_{k}(T) + \frac{2/\pi}{T_{k}-T_{k-1}} \left(M(T_{k-1})  
- M(T_{k}) \right), 
\label{eqn:probab} 
\end{equation} 
%
where we can determine P$_{k}^{i}(E)$ at the $i^{th}$-iteration making
use of the form of P$_{k}^{i-1}(E)$ from the $(i-1)^{th}$-iteration to
compute the correction term $[M(T_{k-1}) - M(T_{k})]$. As an initial
guess, we consider $P_{k}(E)$ equal to $P_{k}(T)$ at a zero-order
approximation. This iterative process is repeated until convergence is
achieved.
 
From the data $P(T)$ of the histogram shown in Fig. \ref{fig:PT}, we
have obtained a satisfactory solution $P(E)$ with a few number (5-10)
of iterations. We will discuss later the physical quality of this
solution, but we must recognize here the stability of the
method. Actually, we have always achieved the same solution with a few
number of iterations, starting from any trial initial distribution
(namely, any zero-order approximation for $P_{k}(E)$). Moreover, the
greater advantage of solving the integral equation by means of an
iterative perturbation method is the possibility to recover, according
to Eq. \ref{eqn:probabi}, the approximate diagram $P_{k}^{i}(T)$ that
corresponds to any iterative solution $P_{k}^{i}(E)$, and this yields
a double check on the evolution and quality of results.
 
However, when dealing with this kind of inverse problem, it may often
happen in the practice that the results are correct mathematically,
but not from the physical standpoint. In view of this difficulty, and
in spite of the excellent quality of the foregoing iterative method,
we wished to develop an alternative method of inversion for the
integral Eq. \ref{eqn:probability1}, in order to double check the
results.
 
The other way to numerically treat the integral Eq. 
\ref{eqn:probability1} comes from the analytical inversion 
 
\begin{equation} 
\int_{T'}^{1}P(E)=\int_{T'}^{1}\frac{T}{\sqrt{T^{2} - T'^{2}}}\,P(T)\,dT'. 
\label{eqn:abelnew} 
\end{equation} 
 
Once we have this analytical form for the required solution, we
rewrite it for the histogram representation of $P(E)$, by defining
$P_k(E)$ as $P_k(T)$ in Eq. \ref{eqn:probabiliE}. Again the difference
between the two integrals of $P(T)$ show up. Now, we impose the
histogram model of this $P(T)$ to derive analytically the matrix
elements that relate any $P_{k}(E)$ to all the elements $P_{k}(T)$.
The corresponding solutions obtained with the two methods are the 
same, allowing for the small differences due to round-off errors of 
the two different numerical algorithms employed. 
 
Now that we are confident of the reliability of both methods of
inversion of the integral equation, we can came back to the aforesaid
difficulties.
 
When applying either method to the observed distribution $P(T)$, in
form of a histogram with bins $P_k(T)$ as shown in Fig. \ref{fig:PT},
we may obtain non-physical results, these are negative values for some
bins $P_k(E)$. This occurs due to the fact that in the frame of the
adopted histogram representation for $P_k(T)$ and $P_k(E)$ and the
associated numerical algorithm chosen to represent the matrix operator
for the integral Eq. \ref{eqn:probability1}, the measured values
$P_{k}(T)$ cannot be the integral transform of any physical
distribution $P(E)$. However, another set of values $P_{k}(T)$ that
are slightly different from the original, and consequently compatible
with the observations, might satisfy the above requirement; i.e., it
can be the integral transform of some physical $P(E)$ and its inverse
transform will solve our problem.

These considerations claim a statistical regularization process, which
can be achieved by considering the histogram $P(T)$ to be the
statistical mean of 1000 histograms, all of them compatible with the
observations according to Poisson statistics. For each one of the 1000
possible realizations for $P_{k}(T)$, we have obtained the
corresponding $P_{k}(E)$ by means of the inversion of the integral
equation. The non-physical histograms $P_{k}(E)$, i.e., those with
some negative bins, were rejected. From the physical solutions, we
have obtained the mean histogram and the corresponding error bars, as
shown in Fig. \ref{fig:PE}.
The statistical regularization process also allows us to estimate
errors due to the possible lack of statistics in the sample.

\subsection{The probability distribution function of intrinsic 
ellipticities} 
\label{sec:results_ellipticity} 
 
\begin{figure} 
\centering 
\includegraphics[width=8.5cm]{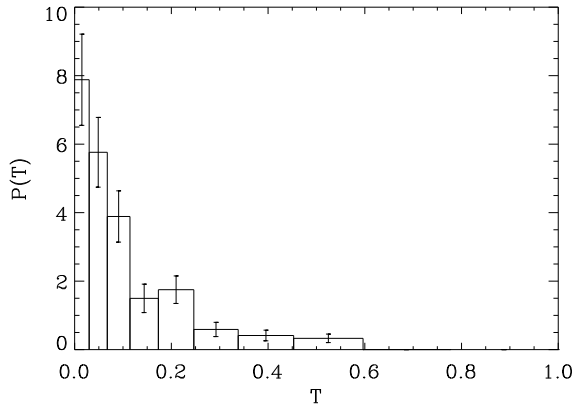} 
\caption{PDF of $T$. The probability is normalized over 10 bins, which
  are geometrically distributed to cover the interval $(0,1)$. The
  width of the first bin is 0.03 and the width ratio of two
  consecutive bins is 1.25.  Error bars correspond to Poisson
  statistics.}
\label{fig:PT} 
\end{figure}

\begin{figure} 
\centering 
\includegraphics[width=8.5cm]{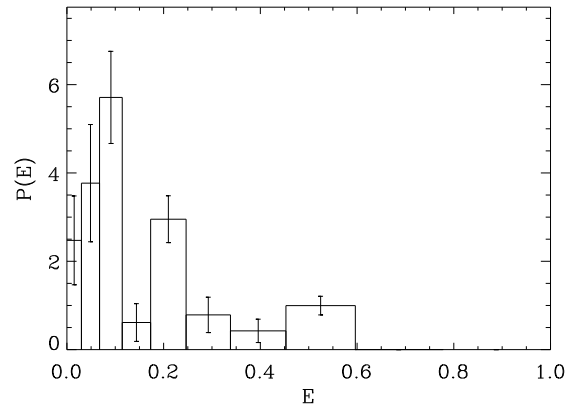} 
\caption{PDF of $E$. The probability is normalized over 10 bins, which
  are geometrically distributed to cover the interval $(0,1)$. The
  width of the first bin is 0.03 and the width ratio of two
  consecutive bins is 1.25.  The error bar of each $P_{k}(E)$ bin
  corresponds to the Poisson statistics of 1000 realizations of
  $P_{k}(T)$ after excluding non-physical cases.}
\label{fig:PE} 
\end{figure}

In Fig. \ref{fig:PE} we present the PDF of the bulge intrinsic
ellipticities $P(E)$. It was obtained by applying the procedure
described in Sect. \ref{sec:methods} using the PDF $P(T)$ shown in
Fig. \ref{fig:PT}. The $T$ values for each galaxy were calculated by
means of Eqs. \ref{eqn:qe} and \ref{eqn:Tdef} from the measured values
of $e$, $\delta$ and $\theta$.

The PDF is characterized by a significant decrease of probability for
$E$$<$$0.07$ (or equivalently $B/A$$<$$0.93$), suggesting that the
shape of bulge ellipsoids in their equatorial plane is most probably
elliptical rather than circular. Such a decrease is caused neither by
the lack of statistics (because in the regularization method we took
into account the Poisson noise) nor by the width of the bins (because we
tried different bin widths).
 
We have calculated the average $E$ value weighted with the PDF through 
 
\begin{equation} 
\langle E \rangle = \frac{\sum_{i}PDF(E_i) * E_i}{\sum_{i}PDF(E_i)}, 
\label{eqn:meanE} 
\end{equation} 
%
obtaining a value of $\langle E \rangle = 0.16$ ($\langle B/A \rangle
= 0.85$). This is fully consistent with previous findings by Bertola
et al. (1991) and Fathi \& Peletier (2003), based on the analysis of
smaller samples of bulges.
For the sake of comparison, the value of $\langle B/A \rangle$ was
derived from their data using Eq. 45. It is $\langle B/A \rangle =
0.85$ for the bulges studied by Bertola et al. (1991). They adopted a
different approach to derive the PDF for intrinsic axial ratio of
bulges from the misalignment of the major axes of bulges and disks and
the apparent ellipticity of bulges. It is $\langle B/A \rangle = 0.79$
for the early-type bulges of the sample studied by Fathi \& Peletier
(2003). They measured the bulge equatorial ellipticity by analyzing
the deprojected ellipticity of the ellipses fitting the galaxy
isophotes within the bulge radius.
%

A further important result derived from $P(E)$ is that there are not
bulges with $E>0.6$ ($B/A<0.5$). This is also in good agreement with
Bertola et al. (1991) and Fathi $\&$ Peletier (2003). They found a
minimum axial ratio $B/A=0.55$ and $B/A=0.45$, respectively.
 
\begin{figure*} 
\centering 
\includegraphics{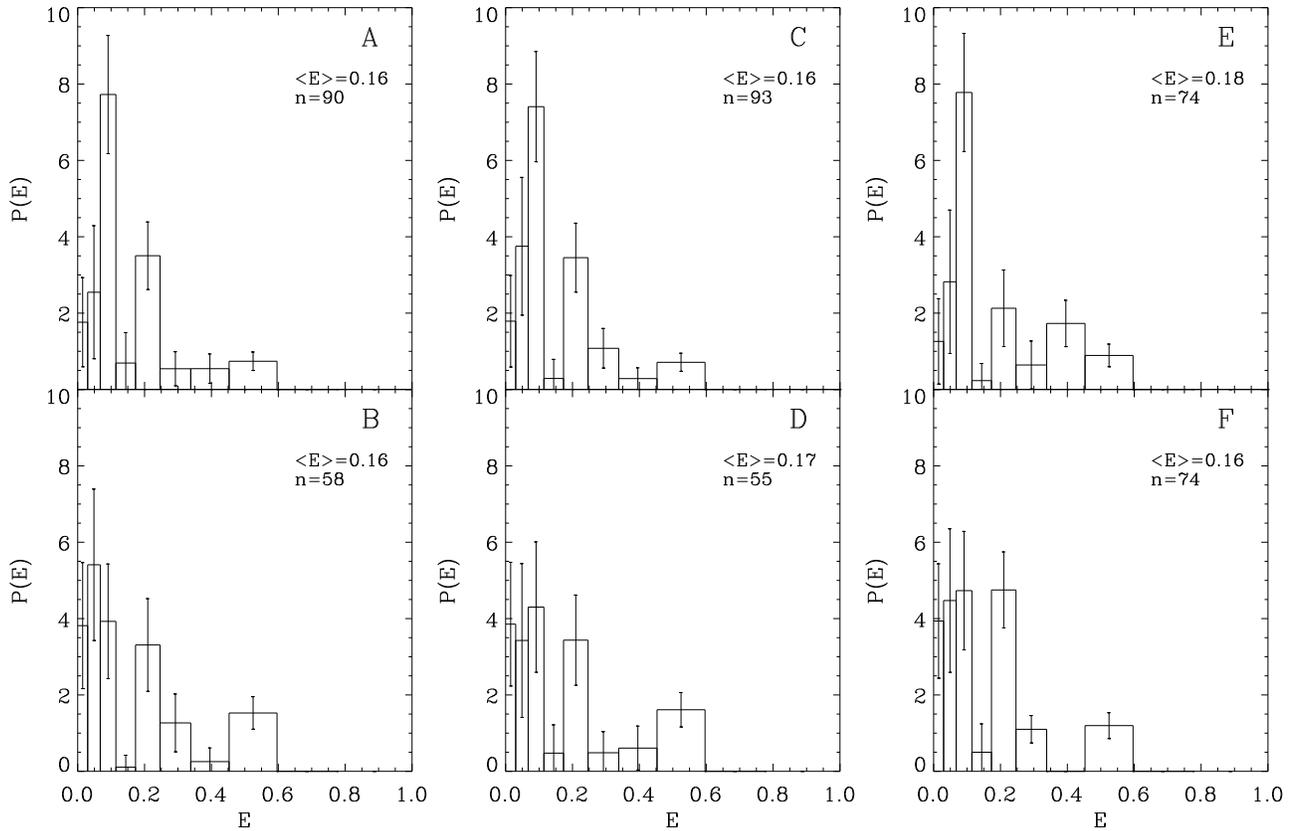} 
\caption{PDF of $E$ for the different subsamples. (A) Lenticular
galaxies ($\rm -3 \leq HT <0$). (B) Spiral galaxies ($\rm 0 \leq HT
\leq 3$). (C) Galaxies with $n\geq2$. (D) Galaxies with $n< 2$. (E)
Faint bulges ($M_{\rm b}\geq-22$). (F) Bright bulges ($M_{\rm
b}<-22$). Bin widths and error bars as in Fig. \ref{fig:PE}. The mean
intrinsic ellipticity and the number of galaxies of the subsample are
given in each panel.}
\label{fig:PEsubsample} 
\end{figure*}

We also studied the possible differences in the shape of bulges 
depending on their observational characteristics (i.e., morphology, 
light concentration, and luminosity, see Fig. \ref{fig:PEsubsample}). 
First of all, we subdivided our bulges according to the morphological
classification $\rm -3 \leq HT <0$ and $\rm 0 \leq HT \leq 3$ of their
host galaxies to look for differences between lenticular and
early-type spiral galaxies. A second test was done by subdividing the
sample bulge between those with a S\'ersic index $n<2$ and those with
$n \ge 2$ to investigate possible correlations of bulge shape with
light concentration. Finally, we subdivided our bulge into faint
($M_{\rm b} \ge -22$) and bright ($M_{\rm b}<-22$) in order to search
for differences of bulge shape with the $J-$band total luminosity.
We did not find any significant difference between the studied
subsamples. They are characterized by the same distribution of $E$, as
confirmed at a high confidence level $(>99\%)$ by a Kolmogorov-Smirnov
test.
 
Several authors discussed the problem of the intrinsic shape of 
elliptical galaxies by means of observations and/or numerical 
simulations. 
Ryden (1992), Lambas et al. (1992), and Bak \& Statler (2000) agree 
that the observed distribution of ellipticities cannot be reproduced 
by any distribution of either prolate or oblate spheroidal 
systems. Any acceptable distribution of triaxial systems is dominated 
by nearly-oblate spheroidal rather than nearly-prolate spheroidal 
systems. 
The formation of triaxial elliptical galaxies via simulation of
merging events in the framework of a hierarchical clustering assembly
was studied by Barnes $\&$ Hernquist (1996), Naab $\&$ Burkert (2003)
and Gonzalez-Garcia $\&$ Balcells (2005). On the other hand, in the
monolithic scenario where the galaxy formation occurs at high redshift
after a rapid collapse, we may expect that the final galaxy shape
would be nearly spherical or axisymmetric, as recently found in
numerical experiments by Merlin \& Chiosi (2006).
But there is no extensive testing of the predictions of numerical 
simulations against the derived distribution of bulge intrinsic 
ellipticities. 
 
%

\subsection{The influence of nuclear bars on $P(E)$} 
 
The presence of nuclear bars in galaxy bulges has been known since the
former work of de Vaucouleurs (1974). However, it is only in the last
decade with the advent of high-resolution imaging that a large number
of them have been detected allowing the study of their demography and
properties \citep[see][and reference therein]{erwin04}.

The sample galaxies were selected to not host large-scale bars,
according to visual inspection and photometric decomposition of their
$J-$band images (Sect. \ref{sec:sample}). These selection criteria did
not account for the presence of nuclear bars.
In fact, our sample has 23 galaxies in common with the samples studied
by Mulchaey \& Regan (1997), Jungwiert et al. (1997), Martini \& Pogge
(1999), Marquez et al. (1999) and Laine et al. (2002). They were
interested in the demography of nuclear bars. A nuclear bar was found
in 6 to 8 out of these 23 galaxies ($26\%$--$35\%$), according to the
different authors' classifications.

Since nuclear bars are more elongated than their host bulges and have
random orientations, they could affect the measurement of the
structural parameters of bulges and consequently their $P(E)$.
To address this issue we carried out a series of simulations on a
large set of artificial galaxies. They were obtained by adding a
nuclear bar to the artificial image of a typical galaxy of the sample
and analyzing the structural properties of the resulting image with
GASP2D, as done in Sect. \ref{sec:model}.

We adopted a Ferrers profile \citep{laurikainen05} to describe the
surface brightness of the nuclear bar component

\begin{displaymath} 
I_{\rm nb}(\xi,\eta) = \left\{\begin{array}{ll}
I_{\rm 0,nb}\left(1-\left(\frac{r_{\rm nb}}{a_{\rm nb}}\right)^2\right)^{n_{\rm nb}+0.5} & 
  {\rm if} \ r_{\rm nb} \leq a_{\rm nb}\\
0 & {\rm if} \ r_{\rm nb} > a_{\rm nb}
\end{array}\right.
\label{eqn:nbar_surfbright} 
\end{displaymath} 
%
where the coordinates $(\xi,\eta)$ are defined as in
Sect. \ref{sec:model} and $a_{\rm nb}$, $I_{\rm 0, nb}$, and $n_{\rm
nb}$ are the bar length, its central surface brightness, and a shape
parameter describing the curvature of the surface-brightness profile,
respectively. Following Laurikainen et al. (2005) we chose $n_{\rm
nb}=2$.
Nuclear bar isophotes are ellipses centered on $(\xi_0,\,\eta_0)$ with
constant position angle PA$_{\rm nb}$ and constant ellipticity
$\epsilon_{\rm nb} = 1 - q_{\rm nb}$.
The radius $r_{\rm nb}$ is given by 
 
\begin{eqnarray} 
r_{\rm nb} & = &\left[(-(\xi-\xi_0) \sin{{\rm PA_{nb}}} + (\eta-\eta_0)  
\cos{{\rm PA_{nb}}})^2 + \right. \nonumber \\  
    &   & - \left.((\xi-\xi_0) \cos{{\rm PA_{nb}}} + (\eta-\eta_0)  
\sin{{\rm PA_{nb}}})^2/q_{nb}^2\right]^{1/2}. 
\label{eqn:nbar_radius} 
\end{eqnarray} 
 
We generated 1000 images of galaxies with a S\`ersic bulge, an
exponential disk, and a Ferrers nuclear bar.
The structural parameters of the bulge and disk were selected to match
those of a typical galaxy of the studied sample. It has $J_T=9.6$ mag,
$r_{\rm e}=0.87$ kpc, $n=2.32$, $h=2.47$ kpc and $B/T=0.37$ according
to the mean values of the structural parameters given in Table
\ref{tab:sample}. Apparent ellipticity and position angle of the bulge
and disk were randomly chosen in the ranges defined in
Sect. \ref{sec:test}.
The structural parameters of the nuclear bar were randomly chosen in
the ranges $0 < a_{\rm nb} < r_{\rm e}$ for the length, $0.2 < q_{\rm
nb} < 0.7$ for the ellipticity, $0^\circ < {\rm PA_{nb}} < 180^\circ$
for the position angle, and $0 < L_{\rm nb}/T < 0.02$ for the nuclear
bar-to-total luminosity ratio. 

The $a_{\rm nb}$ range was estimated from the 5 sample galaxies with a
nuclear bar in common with Laine et al. (2002), which are
characterized by $\langle a_{\rm nb}/r_{\rm e}\rangle=0.8$. Detailed
studies about luminosities of nuclear bars are still
missing. Nevertheless, the $L_{\rm nb}/T$ range was derived by
considering that some nuclear bars are secondary bars, which reside in
large-scale bars. According to Erwin \& Sparke (2002), a typical
secondary bar is about $12\%$ of the size of its primary bar.  From
Wozniak et al. (1995) we derived that the luminosity of the secondary
bar is about $18\%$ of that of the primary one. Since a primary bar
contributes about $15\%$ to the total luminosity of its galaxy
\citep{prieto01, laurikainen05}, the typical $L_{\rm nb}/T$ ratio for
a nuclear bar is about $2\%$.

All simulated galaxies were assumed at a distance of 30 Mpc. Pixel
scale, CCD gain and RON, seeing, background level and photon noise of
the artificial images were assumed as is Sect. \ref{sec:test}. The
two-dimensional parametric decomposition was applied by analyzing with
GASP2D the images of the artificial galaxies as if they were real. We
defined errors on the parameters as the difference between the input
and output values. The mean errors on the fitted ellipticity and
position angle of the bulge and disk and their standard deviation are
given in Table \ref{tab:error_bar}. They correspond to systematic and
typical errors.
 
\begin{table}[h!] 
\caption{Errors on the ellipticity and position angles of the bulge
  and disk calculated for galaxies with nuclear bars by means of Monte
  Carlo simulations.}
\label{tab:error_bar}  
\centering  
\begin{tabular}{c c c} 
\hline\hline 
\noalign{\smallskip} 
Parameter & Mean & St.Dev.\\ 
\hline 
\noalign{\smallskip} 
$\Delta q_{\rm b}$ & 0.02 & 0.03 \\ 
$\Delta q_{\rm d}$ & 0.005 & 0.02 \\ 
$\Delta$PA$_{\rm b}\ (^\circ)$ & 0.3 & 7  \\ 
$\Delta$PA$_{\rm d}\ (^\circ)$ & 0.1 & 3  \\ 
\hline 
\noalign{\bigskip} 
\end{tabular}  
\begin{scriptsize} 
\end{scriptsize} 
\end{table} 
 

For each sample galaxy the values of $e$, $\delta$, and $\theta$ were
derived in Sect. \ref{sec:ellipticity} from the ellipticity and
position angle measured for the bulge and disk.
We randomly generated a series of $q_{\rm b}$, $q_{\rm d}$, ${\rm
PA_{\rm b}}$, and ${\rm PA_{\rm d}}$ by assuming they were normally
distributed with the mean and standard deviation given in Table
\ref{tab:error_bar}.
We tested whether bulges are axisymmetric structures, which appear
elongated and twisted with respect to the disk component due to the
presence of a nuclear bar. To this aim, for each galaxy we selected
1000 realizations of $q_{\rm b}$, $q_{\rm d}$, ${\rm PA_{\rm b}}$, and
${\rm PA_{\rm d}}$ which gave smaller $e$ (i.e., a rounder bulge) and
smaller $\delta$ (i.e., a smaller misalignment between bulge and disk)
with respect to the observed ones. This correction can be considered
as an upper limit to the bulge axisymmetry. We obtained 1000 $P(T)$
distributions and calculated $P(E)$ from their mean.
 
\begin{figure} 
\centering 
\includegraphics[width=8.5cm]{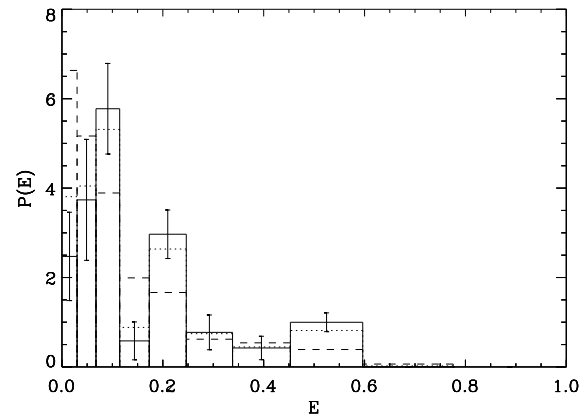} 
\caption{PDF of $E$ for the original sample (solid lines), for a
sample with 30\% of bulges with a nuclear bar (dotted line) and for a
100\% fraction of galaxies hosting a nuclear bar (dashed line). Bin
widths and error bars as in Fig. \ref{fig:PE}.}
\label{fig:PE_comp_bars} 
\end{figure}

Following this procedure, if we consider that all galaxies in our
sample host a nuclear bar, we obtain a $P(E)$ where the decrease of
the probability for $E<0.07$ disappears
(Fig. \ref{fig:PE_comp_bars}). This means that most of the bulges are
circular in the equatorial plane. The average value of the ellipticity
of $\langle E \rangle = 0.12$ ($\langle B/A \rangle = 0.89$). However,
if we consider a more realistic fraction of galaxies that host a
nuclear bar (i.e, 30\% as found by Laine et al. 2002 and Erwin \&
Sparke 2002), the resulting $P(E)$ is consistent within errors with
the $P(E)$ derived in Sect. \ref{sec:results_ellipticity}
(Fig. \ref{fig:PE_comp_bars}). The average value is $\langle E \rangle
= 0.15$ ($\langle B/A \rangle = 0.86$).
 
%
 
\section{Conclusions} 
\label{sec:discussion} 
 
The structural parameters of the bulge and disk of a magnitude-limited 
sample of 148 unbarred S0--Sb galaxies were investigated to constrain 
the dominant mechanism at the epoch of bulge assembly. 
 
\begin{itemize} 
 
\item We presented a new fitting algorithm (GASP2D) to perform
      two-dimensional photometric decomposition of galaxy images. The
      surface-brightness distribution of the galaxy was assumed to be
      the sum of the contribution of a S\'ersic bulge and an
      exponential disk. The two components were characterized by
      elliptical and concentric isophotes with constant (but possibly
      different) ellipticity and position angles.  
      GASP2D is optimized to deal with large image samples, and it
      adopts a robust Levenberg-Marquard fitting algorithm in order to
      obtain reliable estimates of the galaxy structural parameters.
 
\item The bulge and disk parameters of the sample galaxies were
      derived from the $J-$band images, which were available in the
      Two Micron All Sky Survey.
 
\item The bulges of the sample galaxies follow the same FP, FJ, and PP 
      relationships found for elliptical galaxies. No statistically 
      significant difference is observed when only bulges of 
      lenticular and early-to-intermediate spiral galaxies were 
      considered. This supports the idea that bulges and ellipticals 
      formed in the same way. 
 
\item Tight correlations between the parameters of bulges and disks
      were found. In fact, the disk scale lengths increase with both
      the central velocity dispersion and bulge effective
      radius. Therefore, larger disks reside in galaxies with more
      massive and larger bulges. This was interpreted as an indication
      of the formation of bulges via secular evolution of their host
      disks.
 
\item Our measurements of the exponential scale length of the bulge
      and disk, as well as of bulge shape parameter, were also fully
      consistent with numerical simulations of the effects of mergers
      on the mass distribution of the bulge and disk in galaxies
      formed in hierarchical clustering scenarios.
 
\item These results indicate that the above relations are not enough
      to clearly distinguish between bulges formed by early
      dissipative collapse, merging or secular evolution. All these
      mechanisms could be tested against the intrinsic shape of
      bulges. Therefore, the PDF of the intrinsic equatorial
      ellipticity of the bulges was derived from the distribution of
      the observed ellipticities of bulges and their misalignments
      with disks.
 
\item About $80\%$ of bulges in unbarred lenticular and
      early-to-intermediate spiral galaxies are not oblate but
      triaxial ellipsoids. Their mean axial ratio in the equatorial
      plane is $\langle B/A \rangle = 0.85$. This is consistent with
      previous findings by Bertola et al. (1991) and Fathi \& Peletier
      (2003). There is no significant dependence of the PDF on the
      morphology, light concentration, and luminosity of bulges. The derived
      PDF is independent of the possible presence of nuclear bars.
  
\end{itemize} 
 
\begin{acknowledgements}  
 
{JMA acknowledges support from the Istituto Nazionale di Astrofisica 
(INAF).  EMC receives support from grant PRIN2005/32 by Istituto 
Nazionale di Astrofisica (INAF) and from grant CPDA068415/06 by Padua 
University. JALA acknowledges financial support by the Spanish Ministerio de 
Ciencia y Tecnologia grants AYA2004-08260-C03-01. 
ES acknowledges financial support by the Spanish Ministerio de 
Ciencia y Tecnologia grants AYA2004-08260-C03-01 and AYA2004-08243-C03-01, 
as well as by the European Assosciation for Research in Astronomy. 
JMA, EMC and ES acknowledge the Instituto de Astrof\'\i sica 
de Canarias for hospitality while this paper was in progress. 
We wish to thank P.~B. Tissera for kindly provide us the data of their
simulations and A.~Pizzella and L. Morelli for useful discussions.
We also thank the referee, R. F. Peletier, for insightful
comments which helped to improve the original manuscript.
This research has made use of the Lyon Meudon Extragalactic database 
(LEDA) and of the data products from the Two Micron All Sky Survey, 
which is a joint project of the University of Massachusetts and the 
Infrared Processing and Analysis Center/California Institute of 
Technology, funded by the National Aeronautics and Space 
Administration and the National Science Foundation.} 
\end{acknowledgements} 
 
 
\clearpage 
\onecolumn 
\small 
\begin{landscape} 
\begin{longtable}{l l c c c c c c c c l c c c l} 
\caption{\label{tab:sample} Main properties and structural bulge and disk parameters of the galaxies in the sample.}\\ 
\hline\hline 
\multicolumn{1}{c}{Galaxy} & 
\multicolumn{1}{l}{Type}& 
\multicolumn{1}{c}{J$_{T}$}& 
\multicolumn{1}{c}{V$_{3K}$}& 
\multicolumn{1}{c}{D}& 
\multicolumn{1}{c}{$\sigma_0$}& 
\multicolumn{1}{c}{$\mu_{\rm e}$}& 
\multicolumn{1}{c}{$r_{\rm e}$}& 
\multicolumn{1}{c}{$n$}& 
\multicolumn{1}{c}{$q_{\rm b}$}& 
\multicolumn{1}{c}{PA$_{\rm b}$}& 
\multicolumn{1}{c}{$\mu_{0}$}& 
\multicolumn{1}{c}{$h$}& 
\multicolumn{1}{c}{$q_{\rm d}$}& 
\multicolumn{1}{c}{PA$_{\rm d}$}\\ 
\multicolumn{1}{c}{}& 
\multicolumn{1}{l}{}& 
\multicolumn{1}{c}{(mag)}& 
\multicolumn{1}{c}{(km s$^{-1}$)}& 
\multicolumn{1}{c}{(Mpc)}& 
\multicolumn{1}{c}{(km s$^{-1}$)}& 
\multicolumn{1}{c}{(mag arcsec$^{-2}$)}& 
\multicolumn{1}{c}{(arcsec)}& 
\multicolumn{1}{c}{}& 
\multicolumn{1}{c}{}& 
\multicolumn{1}{c}{($^{\circ}$)}& 
\multicolumn{1}{c}{(mag arcsec$^{-2}$)}& 
\multicolumn{1}{c}{(arcsec)}& 
\multicolumn{1}{c}{}& 
\multicolumn{1}{c}{($^{\circ}$)}\\ 
\multicolumn{1}{c}{(1)}& 
\multicolumn{1}{l}{(2)}& 
\multicolumn{1}{c}{(3)}& 
\multicolumn{1}{c}{(4)}& 
\multicolumn{1}{c}{(5)}& 
\multicolumn{1}{c}{(6)}& 
\multicolumn{1}{c}{(7)}& 
\multicolumn{1}{c}{(8)}& 
\multicolumn{1}{c}{(9)}& 
\multicolumn{1}{c}{(10)}& 
\multicolumn{1}{c}{(11)}& 
\multicolumn{1}{c}{(12)}& 
\multicolumn{1}{c}{(13)}& 
\multicolumn{1}{c}{(14)}& 
\multicolumn{1}{c}{(15)}\\ 
\hline 
\endfirsthead 
\caption{continued.}\\ 
\hline\hline 
\multicolumn{1}{c}{Galaxy} & 
\multicolumn{1}{l}{Type}& 
\multicolumn{1}{c}{J$_{T}$}& 
\multicolumn{1}{c}{V$_{3K}$}& 
\multicolumn{1}{c}{D}& 
\multicolumn{1}{c}{$\sigma_0$}& 
\multicolumn{1}{c}{$\mu_{e}$}& 
\multicolumn{1}{c}{r$_{e}$}& 
\multicolumn{1}{c}{$n$}& 
\multicolumn{1}{c}{$q_{b}$}& 
\multicolumn{1}{c}{PA$_{b}$}& 
\multicolumn{1}{c}{$\mu_{0}$}& 
\multicolumn{1}{c}{$h$}& 
\multicolumn{1}{c}{$q_{d}$}& 
\multicolumn{1}{c}{PA$_{d}$}\\ 
\multicolumn{1}{c}{}& 
\multicolumn{1}{l}{}& 
\multicolumn{1}{c}{(mag)}& 
\multicolumn{1}{c}{(km s$^{-1}$)}& 
\multicolumn{1}{c}{(Mpc)}& 
\multicolumn{1}{c}{(km s$^{-1}$)}& 
\multicolumn{1}{c}{(mag arcsec$^{-2}$)}& 
\multicolumn{1}{c}{(arcsec)}& 
\multicolumn{1}{c}{}& 
\multicolumn{1}{c}{}& 
\multicolumn{1}{c}{($^{\circ}$)}& 
\multicolumn{1}{c}{(mag arcsec$^{-2}$)}& 
\multicolumn{1}{c}{(arcsec)}& 
\multicolumn{1}{c}{}& 
\multicolumn{1}{c}{($^{\circ}$)}\\ 
\multicolumn{1}{c}{(1)}& 
\multicolumn{1}{l}{(2)}& 
\multicolumn{1}{c}{(3)}& 
\multicolumn{1}{c}{(4)}& 
\multicolumn{1}{c}{(5)}& 
\multicolumn{1}{c}{(6)}& 
\multicolumn{1}{c}{(7)}& 
\multicolumn{1}{c}{(8)}& 
\multicolumn{1}{c}{(9)}& 
\multicolumn{1}{c}{(10)}& 
\multicolumn{1}{c}{(11)}& 
\multicolumn{1}{c}{(12)}& 
\multicolumn{1}{c}{(13)}& 
\multicolumn{1}{c}{(14)}& 
\multicolumn{1}{c}{(15)}\\ 
\hline 
\endhead 
\hline 
\endfoot 
ESO443-024 &   SA(s)0$^{-}$	        & 9.48 & 5397 & 72.0      & 273 &17.90 & 8.7  & 2.83& 0.64 & 169.9 & 17.99 & 13.8  & 0.90 & 146.0  \\ 
ESO507-025 &   SA0$^{-}$	        & 9.27 & 3602 & 48.0      & 260 &16.97 & 5.5  & 2.08& 0.68 & 95.6  & 17.49 & 13.4  & 0.78 & 87.8   \\ 
ESO445-002 &   SA0$^{-}$      	        & 9.46 & 4528 & 60.4      &  -  &17.04 & 3.9  & 1.80& 0.93 & 84.5  & 18.27 & 18.9  & 0.71 & 102.4  \\ 
IC0750     &   Sab: sp	                & 9.25 & 935  & 17.0${*}$ & 116 &18.24 & 10.2 & 1.57& 0.22 & 42.7  & 16.81 & 11.1  & 0.40 & 36.5   \\ 
IC2035     &   S0 pec		        & 9.81 & 1408 & 16.5${*}$ & 109 &16.30 & 2.8  & 4.13& 0.70 & 35.0  & 16.13 &  5.5  & 0.71 & 91.6   \\ 
IC3152     &   SA0$^{-}$	        & 9.99 & 3600 & 48.0      & 161 &17.80 & 5.1  & 3.06& 0.94 & 62.6  & 17.67 &  9.2  & 0.83 & 43.6   \\ 
IC4310     &   S0 pec		        & 9.81 & 2714 & 36.2      &  -  &17.90 & 3.7  & 2.10& 0.77 & 64.3  & 18.01 & 14.2  & 0.40 & 72.5   \\ 
IC4991	   &   SA(r)0$^{\circ}$ pec     & 9.45 & 5624 & 75.0      &  -  &17.79 & 6.0  & 1.59& 0.79 & 136.5 & 18.10 & 17.2  & 0.73 & 121.2  \\ 
IC5063	   &   SA(s)0$^{+}$:	        & 9.70 & 3183 & 42.4      & 161 &16.77 & 3.2  & 1.46& 0.67 & 118.9 & 17.01 &  9.8  & 0.85 & 117.0  \\ 
IC5267     &   SA(s)0/a                 & 8.29 & 1484 & 19.8      &  -  &16.74 & 5.1  & 1.69& 0.93 & 144.9 & 17.35 & 21.3  & 0.79 & 142.2  \\ 
MCG-02-33-017 & (R)SA(r)0$^{+}$ pec     & 9.69 & 4247 & 56.6      &  -  &17.47 & 4.8  & 2.34& 0.90 & 24.1  & 18.04 & 15.2  & 0.95 & 39.9   \\ 
NGC50      &   S0$^{-}$ pec	        & 9.65 & 5156 & 68.7      &  -  &17.44 & 5.3  & 2.48& 0.64 & 164.1 & 17.96 & 13.8  & 0.68 & 161.9  \\ 
NGC80      &   SA0$^{-}$:	        & 9.92 & 5409 & 72.1      & 260 &16.62 & 2.3  & 1.56& 0.89 & 34.8  & 17.40 &  9.9  & 0.86 &  0.2   \\ 
NGC148     &   S0$^{\circ}$:sp          & 9.99 & 1253 & 18.4${*}$ &  -  &19.28 & 8.5  & 4.21& 0.61 & 89.7  & 19.46 & 16.0  & 0.24 & 88.9   \\ 
NGC254     &   (R)SA(r)0$^{+}$:         & 9.61 & 1358 & 17.1${*}$ &  -  &16.60 & 2.5  & 1.69& 0.82 & 131.8 & 17.87 & 14.5  & 0.50 & 131.8  \\ 
NGC383     &   SA0$^{-}$	        & 9.49 & 4804 & 64.0      & 278 &17.02 & 5.2  & 1.34& 0.87 & 143.5 & 18.18 & 17.4  & 0.89 & 163.1  \\ 
NGC467	   &   SA(s)0$^{\circ}$ pec     & 9.97 & 5146 & 68.6      & 247 &18.99 & 9.9  & 2.96& 0.97 & 46.3  & 18.83 & 15.8  & 0.80 & 140.5  \\ 
NGC470     &   SA(rs)b	                & 9.76 & 2063 & 30.5${*}$ & 122 &16.40 & 2.2  & 0.92& 0.54 & 154.3 & 18.21 & 16.6  & 0.59 & 163.2  \\ 
NGC488     &   SA(r)b	                & 7.91 & 1959 & 29.3${*}$ & 199 &16.80 & 4.9  & 1.51& 0.91 & 0.7   & 17.54 & 29.3  & 0.75 & 7.0    \\ 
NGC524	   &   SA(rs)0$^{+}$	        & 8.13 & 2126 & 32.1${*}$ & 252 &17.35 & 9.6  & 2.52& 0.94 & 22.7  & 17.07 & 19.2  & 0.92 & 76.7   \\ 
NGC615     &   SA(rs)b	                & 9.51 & 1568 & 23.1${*}$ &  -  &17.33 & 2.8  & 2.23& 0.75 & 161.7 & 17.84 & 15.1  & 0.32 & 161.2  \\ 
NGC665     &   (R)S0$^{\circ}$ ?        & 9.85 & 5162 & 68.8      & 189 &17.26 & 3.8  & 1.12& 0.93 & 34.8  & 17.96 & 12.8  & 0.56 & 116.2  \\ 
NGC772     &   SA(s)b	                & 8.17 & 2166 & 32.6${*}$ & 128 &20.83 & 44.2 & 4.93& 0.80 & 119.6 & 18.78 & 42.3  & 0.70 & 125.4  \\ 
NGC883     &   SA(s)0$^{-}$:	        & 9.85 & 5189 & 69.2      & 311 &16.65 & 3.5  & 1.65& 0.80 & 69.6  & 17.35 &  9.5  & 0.93 & 63.9   \\ 
NGC897     &   SA(rs)a                  & 9.71 & 4582 & 61.1      &  -  &17.86 & 4.1  & 3.37& 0.75 & 23.3  & 17.85 & 13.4  & 0.67 & 28.5   \\ 
NGC1024    &  (R')SA(r)ab               & 9.75 & 3304 & 44.1      & 172 &16.73 & 3.7  & 0.93& 0.51 & 164.2 & 17.84 & 13.6  & 0.49 & 144.3  \\ 
NGC1045    &   SA0$^{-}$ pec		& 9.82 & 4404 & 58.7      &  -  &17.71 & 6.2  & 3.25& 0.62 & 58.5  & 18.07 & 11.8  & 0.69 & 58.8   \\ 
NGC1070    &   Sb	                & 9.64 & 3864 & 51.5      &  -  &18.81 & 13.9 & 2.03& 0.81 & 6.6   & 19.24 & 16.8  & 0.79 & 179.0  \\ 
NGC1107    &   S0			& 9.51 & 3225 & 43.0      & 251 &17.45 & 5.2  & 1.81& 0.57 & 143.2 & 17.93 & 15.1  & 0.53 & 137.0  \\ 
NGC1200    &   SA(s)0$^{-}$		& 9.57 & 3866 & 51.5      &  -  &17.34 & 4.6  & 1.96& 0.84 & 44.7  & 18.24 & 17.5  & 0.82 & 94.2   \\ 
NGC1201    &   SA(r)0$^{\circ}$:	& 8.55 & 1524 & 20.2${*}$ & 167 &17.64 & 7.5  & 2.77& 0.74 & 13.0  & 18.53 & 28.7  & 0.43 & 8.5    \\ 
NGC1351	   &   SA0- pec 		& 9.60 & 1426 & 16.9${*}$ & 137 &17.66 & 5.9  & 3.22& 0.62 & 140.8 & 18.27 & 15.2  & 0.59 & 142.4  \\ 
NGC1357    &   SA(s)ab	                & 9.29 & 1882 & 24.7${*}$ & 124 &17.03 & 5.8  & 1.86& 0.94 & 141.1 & 17.55 & 22.7  & 0.75 & 144.3  \\ 
NGC1366    &   S0$^{\circ}$	        & 9.91 & 1195 & 11.6${*}$ & 120 &17.93 & 3.5  & 5.32& 0.81 & 3.5   & 18.12 & 12.2  & 0.31 & 2.6    \\ 
NGC1400	   &   SAO$^{-}$	        & 8.75 & 418  & 5.0${*}$  & 255 &17.00 & 7.4  & 2.95& 0.88 & 27.2  & 17.89 & 17.8  & 0.87 & 56.3   \\ 
NGC1411	   &   SA(r)0$^{-}$:	        & 9.02 & 890  & 11.2${*}$ &  -  &16.48 & 4.3  & 2.69& 0.85 & 16.8  & 17.86 & 17.9  & 0.73 & 12.8   \\ 
NGC1425    &   SA(s)b	                & 9.13 & 1406 & 17.4${*}$ &  -  &18.67 & 5.1  & 3.28& 0.77 & 125.7 & 18.25 & 21.2  & 0.45 & 131.8  \\ 
NGC1482    &   SA0$^{+}$ pec sp         & 9.72 & 1757 & 19.6${*}$ &  -  &16.42 & 4.8  & 1.02& 0.40 & 106.3 & 17.95 & 12.4  & 0.56 & 105.6  \\  
NGC1527    &   SA(r)0$^{-}$:	        & 8.55 & 1006 & 11.0${*}$ & 165 &17.70 & 6.4  & 2.49& 0.79 & 83.0  & 18.01 & 24.7  & 0.39 & 78.3   \\  
NGC1546	   &   SA0$^{+}$	        & 9.08 & 1131 & 13.4${*}$ &  -  &18.03 & 14.7 & 1.04& 0.32 & 148.6 & 18.02 & 17.8  & 0.61 & 145.4  \\ 
NGC1550    &   SA(s)0$^{-}$:		& 9.78 & 3598 & 48.0      & 336 &16.73 & 4.3  & 2.15& 0.79 & 34.1  & 17.91 & 12.1  & 0.96 & 26.2   \\ 
NGC1553	   &   SA(r)0$^{\circ}$ 	& 7.18 & 1098 & 13.4${*}$ & 177 &16.18 & 5.9  & 1.63& 0.67 & 158.1 & 16.31 & 23.4  & 0.63 & 151.3  \\ 
NGC1947    &   S0$^{-}$ pec	        & 8.43 & 1262 & 12.1${*}$ & 134 &18.21 & 12.8 & 2.68& 0.91 & 49.7  & 17.26 & 17.2  & 0.95 & 4.3    \\ 
NGC1956    &   SA(s)a	                & 9.77 & 4881 & 65.1      &  -  &18.46 & 3.9  & 2.33& 0.86 & 69.2  & 18.09 & 15.2  & 0.46 & 67.4   \\ 
NGC2460    &   SA(s)a	                & 9.48 & 1532 & 23.6${*}$ &  -  &18.32 & 7.9  & 2.63& 0.79 & 5.6   & 17.31 & 11.0  & 0.68 & 38.9   \\ 
NGC2629    &   SA(r)0$^{\circ}$:        & 9.81 & 3762 & 50.2      & 298 &16.74 & 3.9  & 2.46& 0.68 & 93.3  & 17.85 & 10.3  & 0.71 & 98.9   \\ 
NGC2639    &   (R)SA(r)a:               & 9.39 & 3376 & 45.0      & 198 &19.01 & 6.2  & 5.35& 0.69 & 130.7 & 16.73 & 9.9   & 0.49 & 135.5  \\  
NGC2768    &   S0		        & 7.93 & 1525 & 23.7${*}$ & 182 &19.27 & 19.8 & 2.80& 0.66 & 96.8  & 18.79 & 42.4  & 0.36 & 93.2   \\ 
NGC2775    &   SA(r)ab	                & 7.92 & 1654 & 17.0${*}$ & 176 &18.06 & 12.4 & 3.06& 0.90 & 158.9 & 17.58 & 30.0  & 0.78 & 160.7  \\ 
NGC2841    &   SA(r)b:	                & 7.01 & 805  & 12.2${*}$ & 205 &19.44 & 28.4 & 3.91& 0.66 & 156.8 & 19.37 & 113.1 & 0.28 & 150.1  \\ 
NGC2911	   &   SA(s)0: pec	        & 9.61 & 3529 & 47.0      & 234 &18.85 & 10.2 & 3.94& 0.78 & 145.2 & 19.33 & 23.7  & 0.69 & 120.9  \\ 
NGC2985    &   (R')SA(rs)ab             & 8.31 & 1386 & 22.4${*}$ & 140 &17.94 & 13.2 & 2.92& 0.82 & 178.4 & 18.22 & 25.8  & 0.87 & 1.5    \\ 
NGC3031    &   SA(s)ab	                & 4.76 & 46   & 1.4 ${*}$ & 161 &16.64 & 39.8 & 3.69& 0.70 & 143.7 & 16.01 & 78.6  & 0.66 & 157.6  \\ 
NGC3065    &   SA(r)0$^{\circ}$         & 9.97 & 993  & 31.3${*}$ & 160 &16.89 & 3.9  & 2.86& 0.97 & 111.4 & 18.89 & 16.4  & 0.92 & 139.4  \\ 
NGC3169    &   SA(s)a pec               & 8.25 & 1581 & 19.7${*}$ & 165 &16.92 & 8.7  & 2.45& 0.69 & 49.2  & 18.23 & 27.7  & 0.66 & 53.7   \\  
NGC3230    &   S0		        & 9.82 & 3141 & 41.9      &  -  &18.03 & 4.2  & 2.65& 0.79 & 137.0 & 18.02 & 14.1  & 0.40 & 116.5  \\ 
NGC3245	   &   SA(r)0$^{\circ}$:	& 8.79 & 1649 & 22.2${*}$ & 210 &15.97 & 3.3  & 1.71& 0.65 & 176.5 & 17.44 & 17.6  & 0.52 & 177.2  \\ 
NGC3277    &   SA(r)ab	                & 9.83 & 1707 & 25.0${*}$ & 205 &18.23 & 7.5  & 3.54& 0.86 & 10.6  & 18.17 & 12.5  & 0.81 & 38.6   \\  
NGC3497    &   SA(s)0$^{\circ}$:	& 9.65 & 4055 & 54.1      & 270 &17.47 & 4.8  & 2.67& 0.76 & 54.8  & 17.94 & 14.3  & 0.64 & 53.0   \\ 
NGC3593    &   SA(s)0/a                 & 8.41 & 970  & 5.5 ${*}$ & 54  &17.67 & 11.0 & 1.18& 0.40 & 89.4  & 18.53 & 33.1  & 0.38 & 89.2   \\ 
NGC3607	   &   SA(s)0$^{\circ}$:	& 7.61 & 1282 & 19.9${*}$ & 223 &16.42 & 8.1  & 1.86& 0.76 & 124.2 & 17.21 & 21.5  & 0.88 & 124.4  \\ 
NGC3619	   &   (R)SA(s)0$^{+}$: 	& 9.57 & 1703 & 27.9${*}$ &  -  &16.31 & 2.8  & 2.04& 0.90 & 50.4  & 17.77 & 13.3  & 0.82 & 47.5   \\ 
NGC3626	   &   (R)SA(rs)0$^{+}$         & 9.10 & 1803 & 26.3${*}$ &  -  &15.64 & 2.5  & 1.98& 0.63 & 165.3 & 17.59 & 16.4  & 0.58 & 167.3  \\ 
NGC3665	   &   SA(s)0$^{\circ}$         & 8.62 & 2324 & 32.4${*}$ & 184 &17.05 & 6.3  & 1.93& 0.75 & 24.5  & 17.20 & 17.1  & 0.78 & 28.2   \\  
NGC3675    &   SA(s)b	                & 7.84 & 1003 & 12.8${*}$ & 106 &17.87 & 8.7  & 2.26& 0.55 & 2.0   & 18.00 & 39.7  & 0.44 & 0.3    \\ 
NGC3801    &   S0/a	                & 9.87 & 3653 & 48.7      & 182 &18.53 & 5.7  & 2.18& 0.86 & 125.9 & 18.33 & 16.0  & 0.51 & 119.2  \\  
NGC3813    &   SA(rs)b:                 & 9.81 & 1729 & 26.4${*}$ &  -  &20.61 & 8.9  & 1.82& 0.78 & 151.4 & 18.27 & 20.5  & 0.36 & 81.4   \\  
NGC3898    &   SA(s)ab	                & 8.58 & 1340 & 21.9${*}$ & 207 &18.13 & 11.9 & 3.75& 0.64 & 107.9 & 19.07 & 29.2  & 0.50 & 106.9  \\  
NGC3900	   &   SA(r)0$^{+}$	        & 9.62 & 2101 & 29.4${*}$ & 144 &18.46 & 5.1  & 3.81& 0.90 & 167.7 & 18.10 & 16.4  & 0.55 & 1.5    \\ 
NGC3998	   &   SA(r)0$^{\circ}$         & 8.33 & 1207 & 21.6${*}$ & 305 &15.20 & 3.8  & 1.59& 0.82 & 133.5 & 16.97 & 16.3  & 0.81 & 135.2  \\ 
NGC4036    &   S0$^{-}$ 		& 8.47 & 1482 & 24.6${*}$ & 181 &18.83 & 8.9  & 4.01& 0.80 & 69.7  & 17.42 & 20.0  & 0.32 & 81.4   \\ 
NGC4087    &   SA0$^{-}$:		& 9.85 & 3657 & 48.8      & 215 &18.38 & 7.3  & 4.20& 0.89 & 36.1  & 18.57 & 13.3  & 0.73 & 39.1   \\ 
NGC4138	   &   SA(r)0$^{+}$	        & 9.12 & 1105 & 17.0${*}$ & 140 &16.11 & 2.4  & 0.93& 0.73 & 151.0 & 16.99 & 13.9  & 0.61 & 151.0  \\ 
NGC4150	   &   SA(r)0$^{\circ}$         & 9.92 & 492  & 9.7${*}$  & 85  &16.51 & 2.8  & 2.21& 0.80 & 143.9 & 17.82 & 13.6  & 0.61 & 148.1  \\ 
NGC4223	   &   SA(s)0$^{+}$:		& 9.99 & 2574 & 34.3      &  -  &19.82 & 11.3 & 4.52& 0.54 & 123.1 & 18.94 & 19.2  & 0.53 & 131.5  \\  
NGC4224    &   SA(s)a: sp               & 9.49 & 2934 & 35.1${*}$ &  -  &18.73 & 6.1  & 3.49& 0.69 & 56.3  & 18.35 & 19.4  & 0.42 & 54.8   \\  
NGC4233    &   S0$^{\circ}$	        & 9.71 & 2723 & 35.1${*}$ & 220 &18.08 & 4.7  & 1.62& 0.86 & 22.1  & 19.49 & 20.5  & 0.25 & 178.3  \\  
NGC4270    &   S0		        & 9.95 & 2725 & 35.1${*}$ & 154 &18.59 & 3.7  & 3.67& 0.86 & 111.9 & 17.74 & 12.1  & 0.40 & 106.0  \\  
NGC4281    &   S0$^{+}$:sp	        & 8.89 & 3072 & 35.1${*}$ & 280 &17.28 & 4.9  & 2.92& 0.58 & 83.8  & 17.72 & 18.2  & 0.41 & 86.6   \\  
NGC4324	   &   SA(r)0$^{+}$	        & 9.42 & 1999 & 35.1${*}$ & 98  &17.80 & 3.9  & 3.31& 0.83 & 68.0  & 17.52 & 13.3  & 0.47 & 51.2   \\ 
NGC4350	   &   SA0		        & 8.76 & 1565 & 16.8${*}$ & 180 &18.87 & 8.1  & 6.00& 0.68 & 30.4  & 17.16 & 14.2  & 0.24 & 27.7   \\ 
NGC4369    &   (R)SA(rs)a               & 9.78 & 1289 & 21.6${*}$ &  -  &17.68 & 6.1  & 1.47& 0.57 & 159.6 & 17.93 & 14.0  & 0.87 & 83.8   \\ 
NGC4377	   &   SA0$^{-}$	        & 9.73 & 1694 & 16.8${*}$ & 144 &15.59 & 2.5  & 1.34& 0.78 & 158.8 & 17.16 & 10.1  & 0.86 & 175.3  \\ 
NGC4378    &   (R)SA(s)a                & 9.42 & 2888 & 35.1${*}$ & 198 &18.01 & 7.2  & 3.83& 0.84 & 164.7 & 17.58 & 11.8  & 0.79 & 162.2  \\ 
NGC4379    &   S0$^{-}$ pec	        & 9.65 & 1390 & 16.8${*}$ & 108 &17.49 & 6.6  & 2.32& 0.73 & 94.8  & 17.85 & 12.5  & 0.86 & 85.5   \\ 
NGC4380    &   SA(rs)b:                 & 9.22 & 1300 & 16.8${*}$ & 62  &18.81 & 6.0  & 1.50& 0.60 & 157.7 & 19.12 & 35.3  & 0.55 & 156.1  \\  
NGC4382	   &   SA(s)0$^{-}$ pec         & 7.06 & 1070 & 16.8${*}$ & 179 &18.06 & 18.7 & 3.39& 0.79 & 36.8  & 18.01 & 53.4  & 0.66 & 10.9   \\ 
NGC4429	   &   SA(r)0$^{+}$	        & 7.73 & 1459 & 16.8${*}$ & 192 &18.01 & 10.7 & 2.05& 0.68 & 90.3  & 18.24 & 43.5  & 0.38 & 98.3   \\ 
NGC4438    &   SA(s)0/a pec             & 8.25 & 395  & 16.8 ${*}$&  -  &17.45 & 6.0  & 2.69& 0.80 & 33.8  & 16.95 & 17.3  & 0.47 & 19.3   \\ 
NGC4450    &   SA(s)ab	                & 7.94 & 2271 & 16.8${*}$ & 129 &17.75 & 6.7  & 3.08& 0.79 & 1.1   & 17.74 & 29.4  & 0.56 & 178.8  \\ 
NGC4459	   &   SA(r)0$^{+}$	        & 8.10 & 1542 & 16.8${*}$ & 171 &18.05 & 14.5 & 3.38& 0.89 & 94.2  & 18.48 & 28.7  & 0.66 & 109.0  \\ 
NGC4492    &   SA(s)a	                & 9.87 & 2101 & 16.8 ${*}$&  -  &19.73 & 10.7 & 4.80& 0.91 & 169.1 & 18.64 & 17.7  & 0.83 & 52.9   \\  
NGC4501    &   SA(rs)b	                & 7.21 & 2602 & 16.8 ${*}$& 161 &18.43 & 11.5 & 3.45& 0.71 & 135.5 & 17.55 & 43.3  & 0.48 & 142.8  \\ 
NGC4528    &   S0$^{\circ}$:	        & 9.87 & 1692 & 22.6      & 117 &17.63 & 3.7  & 3.03& 0.85 & 75.8  & 17.25 &  9.1  & 0.48 & 4.8    \\ 
NGC4578	   &   SA(r)0$^{\circ}$:        & 9.34 & 2616 & 16.8${*}$ & 120 &19.05 & 14.2 & 3.66& 0.73 & 31.3  & 20.15 & 36.4  & 0.61 & 34.5   \\  
NGC4639    &   SA(rs)b                  & 9.65 & 1300 & 16.8 ${*}$& 87  &18.03 & 7.7  & 1.73& 0.69 & 159.9 & 18.53 & 18.3  & 0.87 & 78.7   \\ 
NGC4698    &   SA(s)ab	                & 8.40 & 1328 & 16.8 ${*}$& 133 &19.12 & 13.8 & 4.07& 0.82 & 74.8  & 18.48 & 33.0  & 0.48 & 166.2  \\ 
NGC4736    &   (R)SA(r)ab               & 6.03 & 531  & 4.3${*}$  & 104 &15.55 & 11.4 & 1.81& 0.90 & 17.9  & 16.87 & 40.4  & 0.67 & 96.0   \\ 
NGC4750    &   (R)SA(rs)ab              & 9.04 & 1683 & 26.1${*}$ &  -  &16.76 & 2.1  & 2.32& 0.83 & 53.7  & 16.76 & 12.9  & 0.77 & 128.0  \\ 
NGC4772    &   SA(s)a	                & 9.16 & 1366 & 16.3${*}$ &  -  &19.34 & 12.3 & 3.68& 0.91 & 47.5  & 18.95 & 21.0  & 0.50 & 150.9  \\ 
NGC4789    &   SA0:		        & 9.99 & 8497 & 113.3     & 270 &16.86 & 2.4  & 1.41& 0.79 & 2.4   & 17.64 &  9.6  & 0.69 & 174.0  \\ 
NGC4800    &   SA(rs)b	                & 9.28 & 984  & 15.2${*}$ & 99  &17.29 & 4.1  & 1.97& 0.81 & 50.1  & 16.98 & 12.3  & 0.69 & 21.1   \\ 
NGC4802    &   SA(r)0		        & 9.36 & 1339 & 17.8      &  -  &17.16 & 2.2  & 1.62& 0.85 & 27.9  & 16.93 & 10.9  & 0.71 & 11.1   \\ 
NGC4814    &   SA(s)b	                & 9.99 & 2658 & 39.3${*}$ &  -  &17.29 & 1.7  & 1.01& 0.73 & 102.3 & 17.11 & 9.9   & 0.68 & 115.0  \\ 
NGC4825    &   SA0$^{-}$	        & 9.47 & 4776 & 63.7      &  -  &17.55 & 6.1  & 1.30& 0.69 & 142.5 & 17.86 & 15.5  & 0.68 & 136.0  \\ 
NGC4826    &   (R)SA(rs)ab              & 6.27 & 702  & 4.1${*}$  & 91  &18.09 & 47.0 & 3.60& 0.75 & 150.3 & 17.68 & 114.5 & 0.47 & 148.3  \\ 
NGC5087	   &   SA0:		        & 8.80 & 2121 & 27.8${*}$ & 283 &16.48 & 6.3  & 2.54& 0.52 & 10.8  & 17.19 & 13.3  & 0.65 & 13.6   \\ 
NGC5273	   &   SA(s)0$^{\circ}$         & 9.50 & 1320 & 21.3${*}$ & 66  &17.24 & 3.2  & 2.43& 0.84 & 1.5   & 18.06 & 18.30  & 0.87 & 6.7    \\ 
NGC5292    &   (R')SA(rs)ab             & 9.62 & 4729 & 63.0      &  -  &17.09 & 3.6  & 1.51& 0.65 & 51.3  & 17.82 & 15.1  & 0.81 & 49.3   \\ 
NGC5313    &   Sb	                & 9.80 & 2732 & 37.8${*}$ &  -  &16.15 & 1.5  & 0.62& 0.72 & 33.2  & 17.42 & 12.6  & 0.51 & 45.8   \\ 
NGC5326    &   SAa:	                & 9.80 & 2712 & 37.8${*}$ & 165 &18.17 & 9.1  & 1.65& 0.53 & 133.4 & 18.93 & 22.3  & 0.60 & 133.0  \\ 
NGC5440    &   Sa	                & 9.83 & 3890 & 51.9      &  -  &17.53 & 3.4  & 1.95& 0.70 & 41.7  & 18.42 & 15.7  & 0.36 & 46.1   \\ 
NGC5485	   &   SA0 pec  	        & 9.32 & 2074 & 32.8${*}$ & 159 &18.18 & 8.9  & 2.27& 0.85 & 0.9   & 17.93 & 15.3  & 0.78 & 173.4  \\ 
NGC5533    &   SA(rs)ab                 & 9.75 & 4051 & 54.0      &  -  &17.70 & 5.1  & 2.34& 0.66 & 23.6  & 18.03 & 14.1  & 0.64 & 26.3   \\ 
NGC5614    &   SA(r)ab pec              & 9.50 & 4073 & 54.3      &  -  &17.60 & 6.1  & 3.41& 0.94 & 47.2  & 17.84 & 12.8  & 0.88 & 149.3  \\ 
NGC5631	   &   SA(s)0$^{\circ}$         & 9.39 & 2075 & 32.7${*}$ & 168 &17.29 & 6.1  & 3.61& 0.82 & 125.4 & 17.55 & 11.6  & 0.94 & 157.7  \\ 
NGC5687    &   S0$^{-}$ 		& 9.89 & 2192 & 34.4${*}$ & 190 &18.25 & 7.5  & 2.31& 0.66 & 103.7 & 19.72 & 27.1  & 0.54 & 96.9   \\ 
NGC5838    &   SA0$^{-}$	        & 8.48 & 1555 & 28.5${*}$ & 266 &17.38 & 6.2  & 2.01& 0.82 & 44.9  & 18.65 & 30.6  & 0.34 & 42.8   \\ 
NGC6278    &   S0		        & 9.96 & 2786 & 37.1      & 150 &16.83 & 3.1  & 2.35& 0.66 & 109.7 & 18.08 & 12.6  & 0.53 & 131.4  \\ 
NGC6340    &   SA(s)0/a                 & 9.25 & 1169 & 22.0${*}$ & 144 &17.17 & 4.3  & 3.17& 0.95 & 121.5 & 17.59 & 16.0  & 0.97 & 64.6   \\ 
NGC6851    &   S0		        & 9.69 & 2913 & 38.8      & 224 &17.31 & 5.9  & 3.13& 0.69 & 167.4 & 17.81 & 10.9  & 0.73 & 156.7  \\ 
NGC6861	   &   SA(s)0$^{-}$:	        & 8.66 & 2636 & 35.5${*}$ &  -  &16.67 & 9.1  & 2.11& 0.53 & 142.4 & 18.01 & 17.8  & 0.70 & 133.8  \\ 
NGC6890    &   SA(rs)ab                 & 9.99 & 2323 & 31.8${*}$ &  -  &18.69 & 5.4  & 3.80& 0.55 & 24.8  & 17.18 & 9.8   & 0.70 & 162.3  \\ 
NGC6893    &   SA(s)$^{\circ}$          & 8.95 & 2996 & 39.9      &  -  &16.86 & 5.6  & 1.85& 0.72 & 2.1   & 18.74 & 26.4  & 0.57 & 10.9   \\ 
NGC6920	   &   SA(rs)0$^{\circ}$:       & 9.39 & 2762 & 36.8      &  -  &16.54 & 3.9  & 2.46& 0.82 & 140.4 & 17.59 & 13.1  & 0.84 & 130.2  \\ 
NGC7007	   &   SA0$^{-}$:	        & 9.91 & 2801 & 37.3${*}$ & 125 &16.78 & 3.1  & 1.86& 0.81 & 133.0 & 17.80 & 11.7  & 0.60 & 1.7    \\ 
NGC7020	   &   (R)SA(r)0$^{+}$          & 9.77 & 2987 & 37.8${*}$ &  -  &17.52 & 2.8  & 2.56& 0.89 & 160.6 & 18.20 & 15.3  & 0.40 & 163.9  \\ 
NGC7049	   &   SA(s)0		        & 8.14 & 2051 & 27.6${*}$ & 240 &16.92 & 8.4  & 2.17& 0.74 & 59.9  & 17.04 & 18.1  & 0.76 & 55.8   \\ 
NGC7083	   &   SA(s)bc  	        & 9.37 & 2999 & 38.7${*}$ & 71  &18.19 & 3.6  & 1.34& 0.95 & 16.1  & 17.88 & 18.6  & 0.56 & 7.9    \\ 
NGC7096    &   SA(s)a	                & 9.84 & 2847 & 36.7${*}$ & 248 &18.95 & 12.5 & 4.05& 0.90 & 124.8 & 18.58 & 10.8  & 0.82 & 101.3  \\ 
NGC7135	   &   SA0$^{-}$ pec		& 9.72 & 1799 & 34.7${*}$ &  -  &18.84 & 8.1  & 4.69& 0.74 & 6.6   & 18.90 & 21.1  & 0.64 & 48.8   \\ 
NGC7166    &   SA0$^{-}$		& 9.47 & 2233 & 30.2${*}$ &  -  &16.12 & 2.7  & 1.70& 0.67 & 9.9   & 17.73 & 13.5  & 0.47 & 13.2   \\ 
NGC7172    &   Sa pec	                & 9.44 & 2315 & 33.9${*}$ & 179 &17.35 & 3.3  & 1.16& 0.82 & 80.3  & 17.56 & 15.5  & 0.52 & 95.8   \\ 
NGC7192    &   SA0$^{-}$: 		& 9.37 & 2794 & 31.6${*}$ & 179 &18.08 & 10.0 & 2.91& 0.96 & 8.5   & 18.38 & 16.5  & 0.96 & 137.0  \\ 
NGC7302    &   SA(s)0$^{-}$:	        & 9.99 & 2221 & 33.7${*}$ &  -  &16.31 & 2.2  & 1.35& 0.86 & 88.0  & 17.66 & 11.0  & 0.61 & 99.0   \\ 
NGC7311    &   Sab	                & 9.96 & 4185 & 55.8      &  -  &16.11 & 1.8  & 1.26& 0.82 & 25.8  & 17.14 & 9.3   & 0.51 & 12.2   \\ 
NGC7377	   &   SA(s)0$^{+}$	        & 9.08 & 3036 & 40.5      & 144 &17.93 & 7.7  & 1.78& 0.84 & 105.6 & 18.22 & 21.6  & 0.78 & 102.8  \\ 
NGC7550    &   SA0$^{-}$		& 9.93 & 4789 & 63.8      & 234 &18.68 & 9.2  & 3.40& 0.97 & 179.5 & 18.30 & 12.1  & 0.86 & 146.8  \\ 
NGC7585	   &   (R')SA(s)0$^{+}$ pec     & 9.36 & 3104 & 41.4      & 219 &17.74 & 6.9  & 3.33& 0.69 & 106.4 & 17.70 & 15.1  & 0.82 & 106.4  \\ 
NGC7600    &   S0$^{-}$ sp		& 9.82 & 3086 & 41.1      & 210 &17.96 & 5.9  & 2.75& 0.97 & 175.0 & 17.89 & 11.7  & 0.90 & 150.1  \\ 
NGC7606    &   SA(s)b	                & 8.62 & 1881 & 28.9${*}$ & 147 &17.45 & 2.8  & 1.02& 0.63 & 147.2 & 18.22 & 31.9  & 0.41 & 145.1  \\ 
NGC7625    &   SA(rs)a pec              & 9.90 & 1272 & 23.0${*}$ &  -  &17.77 & 4.3  & 0.89& 0.59 & 58.1  & 16.91 & 9.0   & 0.74 & 32.0   \\ 
NGC7702    &   (R)SA(r)0$^{+}$          & 9.92 & 3062 & 40.8      &  -  &16.64 & 2.6  & 1.02& 0.61 & 99.8  & 17.82 & 12.9  & 0.53 & 114.8  \\ 
NGC7711    &   S0		        & 9.99 & 3705 & 49.4      & 180 &18.62 & 5.7  & 3.30& 0.75 & 93.0  & 19.37 & 19.1  & 0.33 & 9 1.7  \\	   
NGC7722    &   S0/a	                & 9.88 & 3673 & 49.0      & 165 &17.26 & 4.3  & 1.52& 0.61 & 144.8 & 17.37 & 10.5  & 0.85 & 139.8  \\ 
NGC7742    &   SA(r)b	                & 9.60 & 1294 & 22.2${*}$ & 95  &15.66 & 1.4  & 3.10& 0.91 & 12.6  & 16.24 & 7.5   & 0.98 & 180.0  \\ 
NGC7769    &   (R)SA(rs)b               & 9.90 & 3877 & 51.6      &  -  &18.43 & 6.6  & 3.82& 0.60 & 11.6  & 17.59 & 12.1  & 0.87 & 100.4  \\      
UGC12591   &   S0/a	                & 9.99 & 6615 & 88.2      & 288 &16.78 & 3.7  & 1.51& 0.43 & 59.2  & 17.77 & 11.3  & 0.51 & 58.2   \\ 
\end{longtable} 
\begin{minipage}{230mm} 
NOTE. Col. (1): Galaxy name; Col. (2): Morphological classification from  
de Vaucouleurs et al. (1991, hereafter RC3); Col. (3): Total observed J-band 
magnitude from the 2MASS all-sky extended source catalogue (XSC); 
Col. (4): Radial velocity with respect to the CMB radiation from Lyon 
Extragalactic Database (Hereafter LEDA); Col. (5): Distances obtained 
as V$_{3K}$/H$_{0}$ with H$_{0}$=75 km s$^{-1}$Mpc$^{-1}$, distances 
marked with an asterisk have been taken from Tully (1988); Col. (6): 
Central velocity dispersion from LEDA; Col. (7): Effective surface brightness 
 of the bulge; Col. (8): Effective radius of the bulge; Col. (9): Shape  
parameter of the bulge; Col. (10): Axis ratio of the bulge; Col. (11): Position 
 angle of the bulgescale length of the disk; Col. (12): Central surface  
brightness of the disk; Col. (13): Scale length of the disk; Col. (14):  
Axis ratio of the disk; Col. (15): Position angle of the disk. 
\end{minipage} 
\end{landscape}

\end{document}